\documentclass[sn-mathphys,Numbered]{sn-jnl}

\usepackage{graphicx}%
\usepackage{multirow}%
\usepackage{amsmath,amssymb,amsfonts}%
\usepackage{amsthm}%
\usepackage{mathrsfs}%
\usepackage[title]{appendix}%
\usepackage{xcolor}%
\usepackage{textcomp}%
\usepackage{manyfoot}%
\usepackage{booktabs}%
\usepackage{algorithm}%
\usepackage{algorithmicx}%
\usepackage{algpseudocode}%
\usepackage{listings}%

\UseRawInputEncoding
\usepackage{array}
\usepackage{droidsans}
\usepackage{charter}
\usepackage[T1]{fontenc}
\usepackage{setspace}
\usepackage[compact]{titlesec}
\usepackage{hyperref}
\usepackage{CJK}
\usepackage{placeins}
\usepackage{chemformula}
\usepackage{hhline} 
\usepackage{multirow} 
\usepackage{subfig}
\usepackage{float}
\usepackage{siunitx}
\usepackage{mathtools}

\graphicspath{{./Images/}} 

\theoremstyle{thmstyleone}%
\theoremstyle{thmstyletwo}%

\theoremstyle{thmstylethree}%

\begin{document}

\title[Article Title]{A phenomenological model for interstitial hydrogen absorption in niobium}


\author*[1]{\fnm{Arvind} \sur{Ramachandran}}\email{arvindramachandran0522@gmail.com}

\author[2]{\fnm{Houlong} \sur{Zhuang}}\email{hzhuang7@asu.edu}

\author[1]{\fnm{Klaus} \sur{S. Lackner}}\email{Klaus.Lackner@asu.edu}

\affil*[1]{School of Sustainable Engineering and the Built Environment, Arizona State University, Tempe, USA}

\affil[2]{School for Engineering of Matter, Transport and Energy, Arizona State University, Tempe, USA}


\abstract{A phenomenological model has been developed for hydrogen absorption in niobium. The model has 9 free parameters that have a physical basis. The model provides an excellent fit to the highly accurate isotherm data by Veleckis et al. and has been cross validated by limiting the fitting procedure to a training set. The model makes it possible to extract more information from the data than could be extracted with the computational methods available when the measurements were made. The partial molal enthalpy and partial molal entropy of hydrogen dissolution were calculated using the parameter values corresponding to the best fit to the isotherms.
These quantities are consistent with those reported by Veleckis et al. Interpreting the model parameter values reveals insights into the nature of the interaction between hydrogen atoms in niobium. These insights support our previous analysis of hydrogen interactions in niobium using density functional theory calculations.}




\maketitle

\section{Introduction}\label{sec1}

The properties of interstitial hydrogen in group VB metals, niobium, vanadium, and tantalum, have been experimentally studied for several decades. The three most interesting properties established in these studies include the high hydrogen absorption capacity \citep{veleckisThermodynamicPropertiesSystems1969a,albrechtReactionsNiobiumHydrogen1959a,kujiKuji1984Thermodynamic1984,komjathyNiobiumhydrogenSystem1960}, the facile thermal diffusion of hydrogen \citep{schaumanng.Schaumann1970Nb2006,wipfDiffusionCoefficientHeat}, and the electromigration of hydrogen characteristic of an unscreened proton \citep{petersonPeterson1978Nb1978,erckmannErckman1976Nb1976,brouwerBrouwer1989Nb1989,wipfWipf1975Metals1976}. These properties make these metals great candidates in hydrogen separation and transport applications, providing an alternative to traditionally discussed metals like palladium. A detailed account of these unusual properties can be found elsewhere \citep{ramachandranProbingInteractionsInterstitial2020}. \\

This work focuses on understanding hydrogen absorption in group VB metals. Several decades ago, experiments extensively studied hydrogen absorption in group VB metals (niobium, vanadium, tantalum). \citep{veleckisThermodynamicPropertiesSystems1969a,albrechtReactionsNiobiumHydrogen1959a,kujiKuji1984Thermodynamic1984,komjathyNiobiumhydrogenSystem1960} One of the most comprehensive studies on this topic was done by Veleckis et al.\ \cite{veleckisThermodynamicPropertiesSystems1969a}. The experiments reported in their paper measure how much hydrogen these metals can absorb under various experimental conditions. The experiments also reveal important information regarding this physical system, such as the nature of interactions between the \nobreak{hydrogen} atoms in these metals and thermodynamic quantities like the enthalpy and entropy of hydrogen dissolution in these metals. \\

The motivation to revisit these old experimental data from Veleckis et al. is to rethink how modern data analysis techniques and computational tools can be used to analyze the same data and transcend the limitations of the data analysis techniques available at the time. The exceedingly high accuracy of their data warrants this revisit. We believe we can extract more information about this physical system by applying an improved data analysis procedure to these old data. This work focuses on hydrogen absorption in niobium, but the exact procedure can be extended to Veleckis et al.'s data on hydrogen absorption in vanadium and tantalum. Further, the theory and data analysis procedures described here can help us better understand metal hydrogen systems and how to model them. \\ 

We begin this work by reviewing Veleckis et al.'s experiment and investigating their data analysis procedure to calculate thermodynamic quantities, such as the partial molal enthalpy and partial molal entropy of hydrogen dissolution in these metals.\\

\subsection{Veleckis et al.'s Experiment}

An accurately weighed ultra-pure niobium sample was placed in a crucible, which was, in turn, placed inside a high-temperature glass tube. The tube was planted inside a protection tube in contact with a temperature bath. Hydrogen was passed through a cold nitrogen trap, and zirconium turnings in a furnace for purification before introducing it into the high-temperature glass tube. A schematic diagram and more details on the experimental setup can be found in the original article by Veleckis et al. \cite{veleckisThermodynamicPropertiesSystems1969a}. \\

A gas of known pressure, volume, and temperature was introduced into the high-temperature glass tube in a typical absorption run, and the equilibrium pressure was measured. The concentration of hydrogen in the sample was calculated from the pressure difference, the residual volume, and the change in sample weight. Desorption runs were made by removing a measured amount of hydrogen from the tube and measuring the new pressure after allowing the system to equilibrate. The temperature of the bath and the gas introduced were varied to generate a family of absorption isotherms at varying temperatures. The measurements showed no hysteresis effect, indicating that hydrogen absorption in niobium was completely reversible. \\

\subsection{Veleckis et al.'s Experimental Results and Data Analysis}

\begin{figure}[H]
\includegraphics[width=\columnwidth]{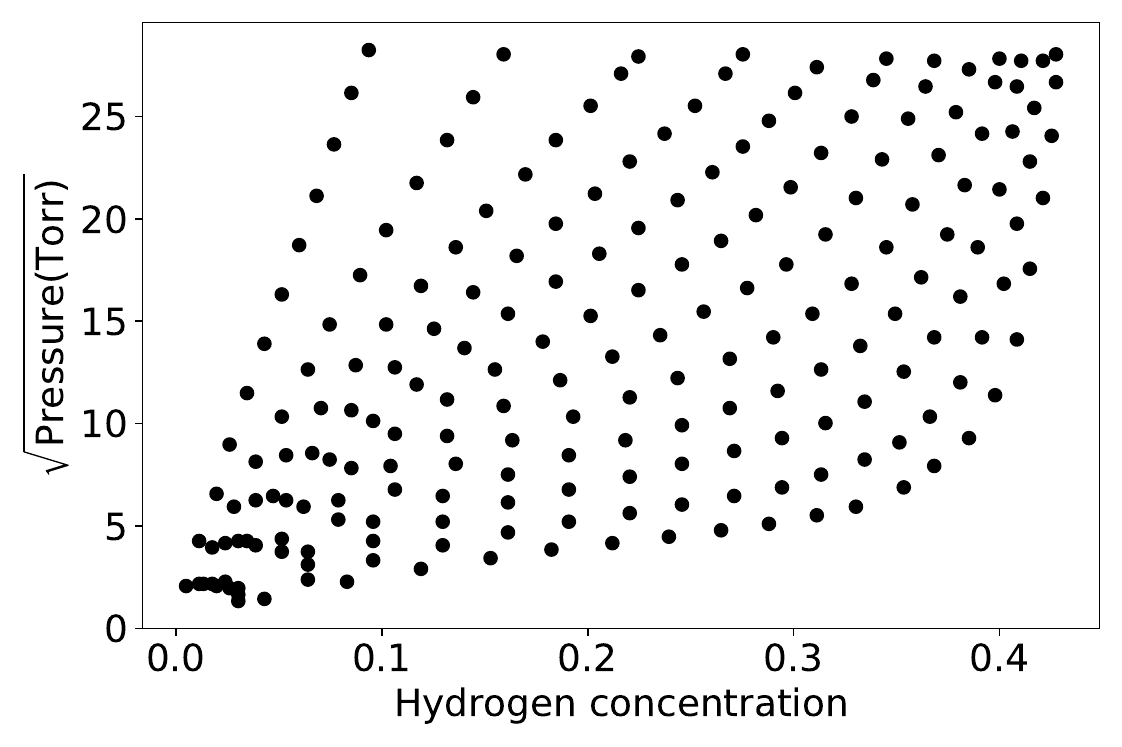}
\caption{Absorption isotherm data digitized from Veleckis et al.\ \cite{veleckisThermodynamicPropertiesSystems1969a}}
\label{fig:4.1}       
\end{figure}

Veleckis et al.\ \cite{veleckisThermodynamicPropertiesSystems1969a} presented the experimental results as absorption isotherms depicting the pressure variation with hydrogen concentration at various temperatures. We used a web-based digitizer \citep{Rohatgi2020} to extract the data from these absorption isotherms. In Fig. \ref{fig:4.1}, we show the digitized data.

Veleckis et al.\ \cite{veleckisThermodynamicPropertiesSystems1969a} expressed the concentration of hydrogen as a mole fraction. However, expressing the hydrogen concentration as the hydrogen-to-metal ratio will be more convenient in the model developed here. The hydrogen to metal ratio ($\mathrm{x}$) is related to the hydrogen mole fraction ($\mathrm{y}$) as 

\begin{equation}
\label{eq:xyrelation}
\mathrm{x} = \frac{\mathrm{y}}{1-\mathrm{y}}
\end{equation}

\begin{figure}[H]
\includegraphics[width=\columnwidth]{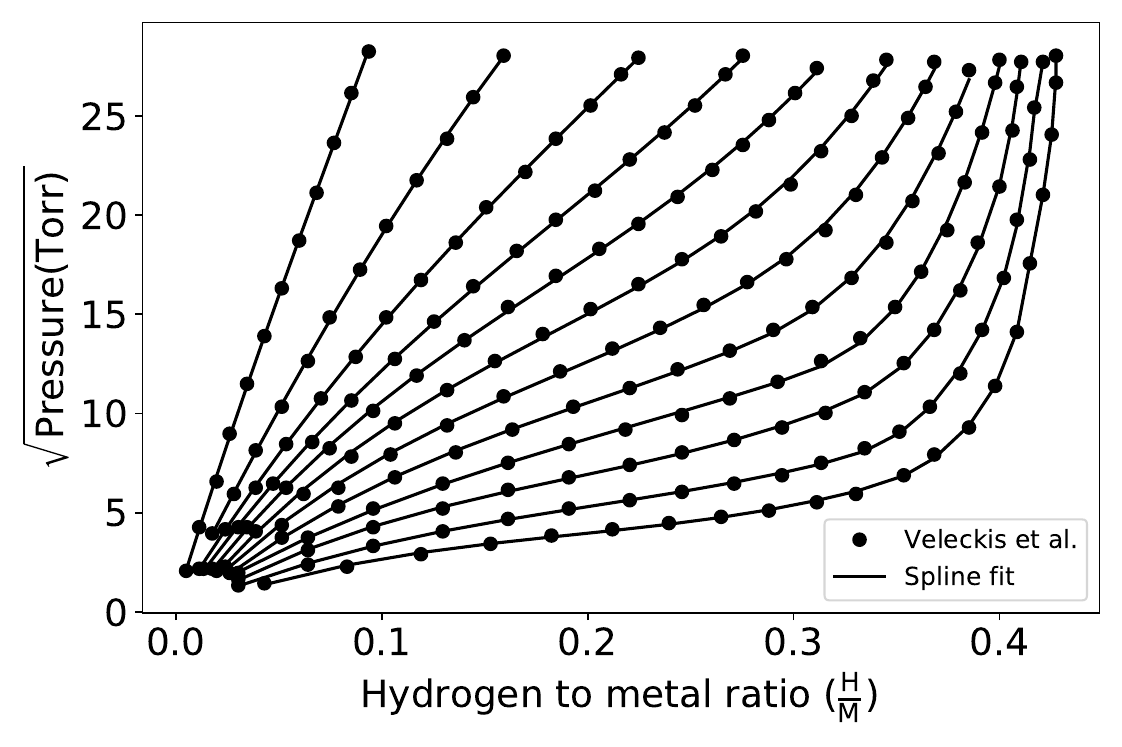}
\caption{Smoothing spline fit to absorption isotherm data digitized from Veleckis et al.\ \cite{veleckisThermodynamicPropertiesSystems1969a}}
\label{fig:4.2}       
\end{figure}

Veleckis et al.\ \cite{veleckisThermodynamicPropertiesSystems1969a} connected the data points by drawing smooth curves through them. We use a 1-D cubic smoothing spline \cite{scipyint60:online} to fit the data points. In Fig. \ref{fig:4.2}, we show the digitized data and our smoothing spline fit of the square root of the pressure ($\mathrm{\sqrt{Pressure (Torr)}}$) as a function of the hydrogen to metal ratio $(\mathrm{\frac{H}{M}})$. 

We provide a summary of the data analysis procedure used by Veleckis et al.\ \cite{veleckisThermodynamicPropertiesSystems1969a} to derive the relative partial molal enthalpy and relative partial molal entropy from these absorption isotherms. \\  

Every data point in Fig. \ref{fig:4.2} corresponds to a chemical equilibrium where the chemical potentials of hydrogen in the gas and metal phases are equal. This \nobreak{relationship is given in Eq. \ref{eq:chemeqbm}}.

\begin{equation}
\label{eq:chemeqbm}
\mu_{\mathrm{H}}^{\mathrm{NbH}} = \frac{1}{2} \mu_\mathrm{H}^{\mathrm{H_2}}
\end{equation}

Where $\mu_{\mathrm{H}}^{\mathrm{NbH}}$ is the chemical potential of hydrogen in the metal phase, and $\mu_\mathrm{H}^{\mathrm{H_2}}$ is the chemical potential of hydrogen in the gas phase. \\

$\mu_{\mathrm{H}}^{\mathrm{NbH}}$ in Eq. \ref{eq:chemeqbm} can be expressed as

\begin{equation}
\label{eq:mu_H^NbH-1}
\mu_\mathrm{H}^{\mathrm{NbH}} = \bar{H}_{\mathrm{H}}^{\mathrm{NbH}} -T\bar{S}_{\mathrm{H}}^{\mathrm{NbH}}
\end{equation}

Where $\bar{H}_{\mathrm{H}}^{\mathrm{NbH}}$ and $\bar{S}_{\mathrm{H}}^{\mathrm{NbH}}$ represent the partial molal enthalpy and partial molal entropy of hydrogen in the metal phase respectively. For a $\mathrm{Nb-H}$ system containing $n_\mathrm{Nb}$ niobium atoms and $n_\mathrm{H}$ atoms, at pressure P and temperature T, the equations below provide the definitions for $\bar{H}_{\mathrm{H}}^{\mathrm{NbH}}$ and $\bar{S}_{\mathrm{H}}^{\mathrm{NbH}}$.

\begin{equation}
\label{eq:Hbar_H}
\bar{H}_{\mathrm{H}}^{\mathrm{NbH}} = \left. \frac{\partial H_{\mathrm{NbH}}}{\partial n_{\mathrm{H}}} \right|_{P, T, n_{\mathrm{Nb}}}
\end{equation}

\begin{equation}
\label{eq:Sbar_H}
\bar{S}_{\mathrm{H}}^{\mathrm{NbH}} = \left. \frac{\partial S_{\mathrm{NbH}}}{\partial n_{\mathrm{H}}} \right|_{P, T, n_{\mathrm{Nb}}}
\end{equation}

In Eq. \ref{eq:Hbar_H} and Eq. \ref{eq:Sbar_H}, $H_{\mathrm{NbH}}$ and $S_{\mathrm{NbH}}$ represent the enthalpy and entropy of the $\mathrm{Nb-H}$ system respectively. \\

$\mu_\mathrm{H}^{\mathrm{H_2}}$ in Eq. \ref{eq:chemeqbm} can be expressed as

\begin{equation}
\label{eq:mu_H^H2-1}
\mu_\mathrm{H}^{\mathrm{H_2}} = {H}_{\mathrm{H_2}}^{\circ} -T{S}_{\mathrm{H_2}}^{\circ} + k_BT\mathrm{\log}(\frac{P}{P_0})
\end{equation}

Where ${H}_{\mathrm{H_2}}^{\circ}$ and ${S}_{\mathrm{H_2}}^{\circ}$ are the standard enthalpy and standard entropy of hydrogen gas respectively. $P_0$ in Eq. \ref{eq:mu_H^H2-1} is 760 Torr or 1 atm and $k_B$ is the boltzmann constant constant. \\

Combining equations \ref{eq:chemeqbm}, \ref{eq:mu_H^NbH-1}, and \ref{eq:mu_H^H2-1}, we get

\begin{equation}
\label{eq:veleckis_eqbm_condition}
\mathrm{\log}(P^{\frac{1}{2}}) = A + \frac{B}{T}
\end{equation}

Where the relative partial molal enthalpy, $\bar{H}_{\mathrm{H}}^{\mathrm{NbH} } - \frac{1}{2} H_{\mathrm{H_2}}^{\circ}$, is related to the value B as 

\begin{equation}
\label{eq:relative_enthalpy_B_relation}
\bar{H}_{\mathrm{H}}^{\mathrm{NbH}} - \frac{1}{2} H_{\mathrm{H_2}}^{\circ} = k_BB
\end{equation}

And the relative partial molal entropy, $\bar{S}_{\mathrm{H}}^{\mathrm{NbH}} - \frac{1}{2} S_{\mathrm{H_2}}^{\circ}$, is related to the value A as 

\begin{equation}
\label{eq:relative_entropy_A_relation}
\bar{S}_{\mathrm{H}}^{\mathrm{NbH}} - \frac{1}{2} S_{\mathrm{H_2}}^{\circ} = k_B(\mathrm{\log}(P_0^{\frac{1}{2}})-A)
\end{equation}

Veleckis et al.\ \cite{veleckisThermodynamicPropertiesSystems1969a} considered a set of mole fractions of hydrogen, $\{\mathrm{y}_i\}$, with $\mathrm{y}_i$ varying from 0 to 0.39, in intervals of 0.01. For each $\mathrm{y}_i$ in this set, they fit the equilibrium pressures at various temperatures corresponding to the different isotherms to the expression in Eq. \ref{eq:veleckis_eqbm_condition} using a least squares treatment. Through this fitting procedure, they derived A and B values at every $\mathrm{y}_i$ value they considered. They calculated $\bar{S}_{\mathrm{H}}^{\mathrm{NbH}} - \frac{1}{2} S_{\mathrm{H_2}}^{\circ}$ and $\bar{H}_{\mathrm{H}}^{\mathrm{NbH}} - \frac{1}{2} H_{\mathrm{H_2}}^{\circ}$ at varying $\{\mathrm{y}_i\}$ from the A and B values. \\

We repeated Veleckis et al.'s fitting procedure to re-derive the relative partial molal enthalpy and the relative partial molal entropy at varying hydrogen concentrations. The reason for repeating this fitting procedure is threefold. Firstly, we want to examine the goodness of fitting the equilibrium pressures at various concentrations to the expression in Eq. \ref{eq:veleckis_eqbm_condition}. Secondly, we want to see if we can reproduce the relative partial molal enthalpy and entropy so we can be sure we correctly understand the physical model and fitting procedure used by Veleckis et al.\ \cite{veleckisThermodynamicPropertiesSystems1969a}. Finally, we want to ensure that our digitizing of the absorption isotherm data did not introduce significant errors. We successfully reproduced Veleckis et al.'s fitting procedure. Fig. S2 and Fig. S3 in the supporting information manuscript show that our estimates for the relative partial molal enthalpy and the relative partial molal entropy were in close agreement with those reported in the original work. The details of this fit and the fitting results can be found in the supporting information manuscript. \\

When examining Veleckis et al.'s data analysis, we identify two key areas for improvement \cite{veleckisThermodynamicPropertiesSystems1969a}. \\

Firstly, for each of the 40 different hydrogen concentrations considered, Veleckis et al.\ \cite{veleckisThermodynamicPropertiesSystems1969a} performed an ordinary least-squares (OLS) to fit the parameterization given in Eq. \ref{eq:veleckis_eqbm_condition}. They obtained the corresponding $A_i$ and $B_i$ values and, in turn, the relative partial molal enthalpy and relative partial molal entropy values. This treatment allows these quantities to vary arbitrarily with hydrogen concentration and does not require them to follow systematic trends. Additionally, this treatment allows for 80 free parameters (40 different concentrations considered, with two free parameters per concentration). This procedure results in a large number of fit parameters, considering there are only 207 data points. Finally, with this approach, if one were to calculate the relative partial molal enthalpy and relative partial molal entropy values with finer granularity in concentration, that would require even more free parameters. There is no upper bound on the number of free parameters required when using their approach. \\

Secondly, Veleckis et al.\ \cite{veleckisThermodynamicPropertiesSystems1969a} assumed that A and B, and by extension, $\bar{S}_{\mathrm{H}}^{\mathrm{NbH}} - \frac{1}{2} S_{\mathrm{H_2}}^{\circ}$ and $\bar{H}_{\mathrm{H}}^{\mathrm{NbH}} - \frac{1}{2} H_{\mathrm{H_2}}^{\circ}$ do not depend on temperature. We expect that the contributions to $\bar{H}_{\mathrm{H}}^{\mathrm{NbH}}$ and $\bar{S}_{\mathrm{H}}^{\mathrm{NbH}}$ from the vibrational degrees of freedom of the interstitial hydrogen atom to be non-negligible in the temperature range 625.65 K -- 944.15 K. Likewise, we also expect the contributions to $H_{\mathrm{H_2}}^{\circ}$ and $S_{\mathrm{H_2}}^{\circ}$ from the translation, vibrational, and rotational degrees of freedom to be non-negligible 
in this temperature range. \\

In this work, we build on Veleckis et al.'s \cite{veleckisThermodynamicPropertiesSystems1969a} treatment by developing a phenomenological model for hydrogen absorption in niobium that describes this physical system with fewer parameters and accounts for the temperature dependencies of enthalpy and entropy in the gas and metal phases. In the next section, we present the development of this phenomenological model \\

\section{A Phenomenological Model for Hydrogen Absorption in Niobium}\label{sec2}

We consider here a niobium-hydrogen system, $\mathrm{NbH_x}$, with a hydrogen to niobium ratio of $\mathrm{x}$. We note that this is a simplified representation of the system. In reality, there is a large number of niobium atoms ($N_{\mathrm{Nb}}$) and a large number of hydrogen atoms ($N_{\mathrm{H}}$) moving freely and largely independently through the niobium lattice. $\mathrm{x}$ is simply the ratio of $N_{\mathrm{H}}$ and $N_{\mathrm{Nb}}$ in such a system. However, the reduced system $\mathrm{NbH_x}$ is sufficient for the purposes model developed here. \\

The expression for chemical equilibrium given by Eq. \ref{eq:chemeqbm} still holds. However, how do we go about deriving $\mu_{\mathrm{H}}^{\mathrm{NbH}}$ and $\frac{1}{2} \mu_\mathrm{H}^{\mathrm{H_2}}$ will be different to the approach taken by Veleckis et al.\ \cite{veleckisThermodynamicPropertiesSystems1969a} We begin by recalling the expression for $\mu_H^{\mathrm{NbH}}$ in Eq. \ref{eq:mu_H^NbH-1}

\begin{equation}
\label{eq:mu_H^NbH-2}
\mu_H^{\mathrm{NbH}} = \bar{H}_{\mathrm{H}}^{\mathrm{NbH}} -T\bar{S}_{\mathrm{H}}^{\mathrm{NbH}}
\end{equation}

In Eq. \ref{eq:mu_H^NbH-2}, $\bar{H}_{\mathrm{H}}^{\mathrm{NbH}}$ and $\bar{S}_{\mathrm{H}}^{\mathrm{NbH}}$ are the partial molal enthalpy and partial molal entropy of hydrogen in the niobium-hydrogen phase respectively. One of the fundamental differences between the model developed here and the treatment by Veleckis et al.\ \cite{veleckisThermodynamicPropertiesSystems1969a} is that we will account for temperature corrections to $\bar{H}_{\mathrm{H}}^{\mathrm{NbH}}$ and $\bar{S}_{\mathrm{H}}^{\mathrm{NbH}}$. \\
 
$\bar{H}_{\mathrm{H}}^{\mathrm{NbH}}$ by definition is related to $H_{\mathrm{NbH_x}}$, the enthalpy of the niobium-hydrogen system, as 

\begin{equation}
\label{eq:Hbar_H^NbH-def}
\bar{H}_{\mathrm{H}}^{\mathrm{NbH}} = \frac{\partial H_{\mathrm{NbH_x}}}{\partial \mathrm{x}}
\end{equation}

Where $H_{\mathrm{NbH_x}}$ can be expressed as 

\begin{equation}
\label{eq:H_NbHx}
H_{\mathrm{NbH_x}} = U_{\mathrm{NbH_x}} + (PV)_{\mathrm{NbH_x}} 
\end{equation}

The first term in Eq. \ref{eq:H_NbHx}, $U_{\mathrm{NbH_x}}$, is the internal energy of the system and the second term, $(PV)_{\mathrm{NbH_x}}$, is the pressure correction term. As seen below, it has three contributions. $E_{\mathrm{NbH_x}}$, the static energy at 0 K, $ZPE_{\mathrm{NbH_x}}$, the zero-point energy corrections, and finally, the temperature corrections from $\mathrm{C_{v,\;NbH_x}}$, which is the heat capacity of the system.

\begin{equation}
  \label{eq:U_NbHx}
U_{\mathrm{NbH_x}} =  E_{\mathrm{NbH_x}}  + ZPE_{{\mathrm{NbH_x}}}  + \int_0^T C_{v,\;\mathrm{NbH_x}} \;\mathrm{d}T
\end{equation}

We assume that $E_{\mathrm{NbH_x}}$ comprises the energy of the niobium atoms, the energy of the hydrogen atoms, and the interaction energy between the hydrogen atoms. In our previous work, it was shown that the interactions between hydrogen atoms are of many-body nature \citep{ramachandranProbingInteractionsInterstitial2020}. We expect the two-particle interaction term to vary like $\mathrm{x^2}$, the three-particle interaction term to vary like $\mathrm{x^3}$, and so on. In Eq. \ref{eq:Ch4-E_NbHx}, we express $E_{\mathrm{NbH_x}}$ to include up to four-particle interactions. $\alpha$, $\beta$, and $\gamma$ in Eq. \ref{eq:Ch4-E_NbHx} are the pre-factors that correspond to the strength of the interactions. 

\begin{equation}
  \label{eq:Ch4-E_NbHx}
E_{\mathrm{NbH_x}} =  E_{\mathrm{Nb}} +  \mathrm{x}E_{\mathrm{H}} + \alpha \mathrm{x}^2 + \beta\mathrm{x}^3 + \gamma\mathrm{x}^4
\end{equation}

Assuming the hydrogen atoms' vibrational modes as independent and treating all hydrogen atoms as identical from a vibrational degree of freedom perspective, we consider that all hydrogen atoms have the same vibrational frequencies. With these assumptions, $ZPE_{{\mathrm{NbH_x}}}$ can be simplified as, 

\begin{equation}
  \label{eq:ZPE_NbHx}
ZPE_{{\mathrm{NbH_x}}} =  \frac{1}{2} \sum_{i=1}^{3} \hbar \nu_{\mathrm{Nb}}^i + \frac{\mathrm{x}}{2} \sum_{i=1}^{3} \hbar \nu_H^i
\end{equation}

Where, $\nu_{\mathrm{Nb}}^i$ for i=1,2,3 are the three vibrational modes of the niobium atom, and $\nu_{\mathrm{H}}^i$ for i=1,2,3 are the three vibrational modes of any hydrogen atom. Our system's niobium and interstitial hydrogen atoms possess only vibrational degrees of freedom. Therefore, the heat capacity can be written as follows. 

\begin{equation}
  \label{eq:Cv_NbHx}
\int_0^T C_{v\;\mathrm{NbH_x}} \;\mathrm{d}T = \int_0^T C_{v,\;\mathrm{Nb}\;vib} \;\mathrm{d}T + \mathrm{x} \int_0^T C_{v,\;\mathrm{H}\;vib} \;\mathrm{d}T 
\end{equation}

In the harmonic approximation for the vibrations of the niobium and hydrogen atoms, their heat capacities can be expressed as,

\begin{equation}
  \label{eq:Cv_Nb}
\int_0^T C_{v,\;\mathrm{Nb}\;vib} \;\mathrm{d}T =  \sum_{i=1}^{3} \frac{\hbar \nu_{\mathrm{Nb}^i}}{e^{\frac{\hbar\nu_{\mathrm{Nb}^i}}{k_BT}}-1}
\end{equation}

\begin{equation}
  \label{eq:Cv_H}
\int_0^T C_{v,\;\mathrm{H}\;vib} \;\mathrm{d}T = \sum_{i=1}^{3} \frac{\hbar \nu_{\mathrm{H}}^i}{e^{\frac{\hbar\nu_{\mathrm{H}^i}}{k_BT}}-1}
\end{equation}

Combining \ref{eq:Cv_NbHx}, \ref{eq:Cv_Nb}, and \ref{eq:Cv_H}, we get

\begin{equation}
  \label{eq:Cv_NbHx-final}
\int_0^T C_{v\;\mathrm{NbH_x}} \;\mathrm{d}T =  \sum_{i=1}^{3} \frac{\hbar \nu_{\mathrm{Nb}}^i}{e^{\frac{\hbar\nu_{\mathrm{Nb}^i}}{k_BT}}-1} +  \mathrm{x}\sum_{i=1}^{3} \frac{\hbar \nu_{\mathrm{H}}^i}{e^{\frac{\hbar\nu_{\mathrm{H}^i}}{k_BT}}-1} 
\end{equation}

Now we present a back-of-the-envelope calculation to show that the contribution from the pressure correction term in Eq. \ref{eq:H_NbHx}, $(PV)_{\mathrm{NbH_x}}$, is insignificant in this model. The metal specimen's external pressure is the pressure of the $\mathrm{H_2}$ gas above it. In Veleckis et al.'s \cite{veleckisThermodynamicPropertiesSystems1969a} experiment, the maximum pressure is less than $10^5$ Pa. The volume of $\mathrm{NbH_x}$ is $\sim 10 \mathrm{\si{\angstrom}^3}$. Therefore the $(PV)_{\mathrm{NbH_x}}$ term is $\sim 10^{-27}$ kJ, which is less than $10^{-5}$ eV. Since the expected enthalpy and entropy contributions are of the order $10^{-2}$ eV, we will ignore the $(PV)_{\mathrm{NbH_x}}$ term in this derivation. \\

From equations \ref{eq:H_NbHx}, \ref{eq:U_NbHx}, \ref{eq:Ch4-E_NbHx}, \ref{eq:ZPE_NbHx}, and \ref{eq:Cv_NbHx-final}, we can express $H_{\mathrm{NbH_x}}$ as 

\begin{equation}
\begin{aligned}
  \label{eq:Ch4-H_NbHx-final}
H_{\mathrm{NbH_x}} =  E_{\mathrm{Nb}} + \frac{1}{2} \sum_{i=1}^{3} \hbar \nu_{\mathrm{Nb}}^i + \sum_{i=1}^{3} \frac{\hbar \nu_{\mathrm{Nb}^i}}{e^{\frac{\hbar\nu_{\mathrm{Nb}^i}}{k_BT}}-1} + \mathrm{x}E_{\mathrm{H}} \\
+ \frac{\mathrm{x}}{2} \sum_{i=1}^{3} \hbar \nu_\mathrm{H}^i + \mathrm{x} \sum_{i=1}^{3} \frac{\hbar \nu_{\mathrm{H}}^i}{e^{\frac{\hbar\nu_{\mathrm{H}^i}}{k_BT}}-1} + \alpha \mathrm{x}^2 + \beta\mathrm{x}^3 + \gamma\mathrm{x}^4
\end{aligned}
\end{equation}

Combining Eq. \ref{eq:Hbar_H^NbH-def} and Eq. \ref{eq:Ch4-H_NbHx-final} we get

\begin{equation}
\begin{aligned}
  \label{eq:Hbar_H^NbHx-final}
\bar{H}_{\mathrm{NbH_x}} = E_{\mathrm{H}} + \frac{1}{2} \sum_{i=1}^{3} \hbar \nu_{\mathrm{H}}^i + \sum_{i=1}^{3} \frac{\hbar \nu_{\mathrm{H}}^i}{e^{\frac{\hbar\nu_{\mathrm{H}^i}}{k_BT}}-1} \\
+ 2\alpha \mathrm{x} + 3\beta\mathrm{x}^2 + 4\gamma\mathrm{x}^3 
\end{aligned}
\end{equation}

$\bar{S}_{\mathrm{H}}^{\mathrm{NbH}}$ in Eq. \ref{eq:mu_H^NbH-2}, by its definition, is related to $S_{\mathrm{NbH_x}}$, the entropy of the niobium-hydrogen system, as follows

\begin{equation}
\begin{multlined}
\label{eq:Sbar_H^NbHx-def}
\bar{S}_{\mathrm{H}}^{\mathrm{NbH}} = \frac{\partial S_{\mathrm{NbH_x}}}{\partial \mathrm{x}}
\end{multlined}
\end{equation}

The two components of $S_{\mathrm{NbH_x}}$ dependent on x are the configurational and non-configurational entropy of the hydrogen atoms. Therefore, $\bar{S}_{\mathrm{H}}^{\mathrm{NbH}}$ is expressed as

\begin{equation}
\label{eq:Sbar_H^NbHx-1}
\bar{S}_{\mathrm{H}}^{\mathrm{NbH}} = \bar{S}_{\mathrm{H}}^{\mathrm{NbH},\ c} + \bar{S}_{\mathrm{H}}^{\mathrm{NbH},\ nc}
\end{equation}

Where $\bar{S}_{\mathrm{H}}^{\mathrm{NbH},\ c}$ and $\bar{S}_{\mathrm{H}}^{\mathrm{NbH},\ nc}$ represent the configurational and non-configurational partial molal entropy of hydrogen in the niobium-hydrogen phase respectively. To derive an expression for $\bar{S}_{\mathrm{H}}^{\mathrm{NbH},\ c}$, we consider the unreduced niobium-hydrogen with $N_{\mathrm{Nb}}$ niobium atoms, $N_{\mathrm{H}}$ hydrogen atoms, and $N_{\mathrm{T}}$ different interstitial tetrahedral sites available for occupation. Applying Boltzmann's equation, $S_{\mathrm{H}}^{\mathrm{NbH},\ c}$ can be related to the number of micro states as 

\begin{equation}
\label{eq:S_Hc^NbHx-1}
S_{\mathrm{H}}^{\mathrm{NbH},\ c} = k_B\mathrm{\log}\biggr[\frac{N_{\mathrm{T}}!}{N_{\mathrm{H}}!(N_{\mathrm{T}}-N_{\mathrm{H}})!}\biggr]
\end{equation}

Applying Stirling's approximation for large $N_{\mathrm{T}}$ and $N_{\mathrm{H}}$

\begin{equation}
\begin{aligned}
\label{eq:S_Hc^NbHx-2}
S_{\mathrm{H}}^{\mathrm{NbH},\ c} = k_B[N_{\mathrm{T}}\mathrm{\log}(N_{\mathrm{T}}) - N_{\mathrm{H}}\mathrm{\log}(N_{\mathrm{H}}) \\
-  (N_{\mathrm{T}}-N_{\mathrm{H}})\mathrm{\log}(N_{\mathrm{T}}-N_{\mathrm{H}})]
\end{aligned}
\end{equation}

\begin{equation}
\begin{aligned}
\label{eq:S_Hc^NbHx-3}
\bar{S}_{\mathrm{H}}^{\mathrm{NbH},\ c} = \frac{\partial S_{\mathrm{H}}^{\mathrm{NbH},\ c}}{\partial N_{\mathrm{H}}} = -k_B[\mathrm{\log}(N_{\mathrm{H}}) - \mathrm{\log}(N_{\mathrm{T}}-N_{\mathrm{H}})]
\end{aligned}
\end{equation}

\begin{equation}
\begin{aligned}
\label{eq:S_Hc^NbHx-4}
\bar{S}_{\mathrm{H}}^{\mathrm{NbH},\ c} = -k_B\mathrm{log}\biggr[\frac{N_{\mathrm{H}}}{N_{\mathrm{T}}-N_{\mathrm{H}}}\biggr]
\end{aligned}
\end{equation}

\begin{equation}
\begin{aligned}
\label{eq:Sbar_Hc^NbHx-final}
\bar{S}_{\mathrm{H}}^{\mathrm{NbH},\ c} = -k_B\mathrm{log}\biggr[\frac{\mathrm{x}}{R-\mathrm{x}}\biggr]
\end{aligned}
\end{equation}

Where $R = \frac{N_{\mathrm{T}}}{N_{\mathrm{Nb}}}$ is the number of interstitial sites accessible per niobium atom and $\mathrm{x} = \frac{N_{\mathrm{H}}}{N_{\mathrm{Nb}}}$ is the hydrogen to metal ratio, we expect $R$ to have a dependence on the temperature. The repulsive interactions between hydrogen atoms occupying neighboring sites will outweigh the entropic contribution to the Gibbs energy from occupying those sites at lower temperatures. However, the entropic contribution dominates at higher temperatures, and $R$ converges to a maximum value of 6 since the host metal, niobium, has a BCC structure. From this qualitative discussion, we suspect $R$ to have a sigmoid-al dependence on temperature and range between a low asymptotic value at 0 K and a high asymptotic \nopagebreak[4] value of 6 at infinite temperature. However, since the range of temperatures of interest here is relatively small (625.65 - 944.15 K), for simplicity, we will assume that $R$ has a quadratic dependence on temperature as in Eq. \ref{eq:R_temp_dep}.

\begin{equation}
\label{eq:R_temp_dep}
  R = R_0 + R_1T + R_2T^2
\end{equation}

The non-configurational entropy of hydrogen in the metal phase, $S_{\mathrm{H}}^{\mathrm{NbH},\ nc}$, arises from the vibrational modes of the interstitial hydrogen atoms and is given by  

\begin{equation}
  \label{eq:S_Hnc^NbHx}
S_{\mathrm{H}}^{\mathrm{NbH},\ nc} = k_B \mathrm{x} \sum_{i=1}^{3} \Biggr[ \frac{\hbar \nu_{\mathrm{H}}^i}{\mbox{\large\(%
k_BT(%
\)}e^{\frac{\hbar\nu_{\mathrm{H}^i}}{k_BT}}-1\mbox{\large\(%
)%
\)}} 
- \mathrm{\log} \Big[1-e^{\frac{-\hbar\nu_{\mathrm{H}^i}}{k_BT}} \Big]   \Biggr]
\end{equation}

And the non-configurational partial molal entropy, $\bar{S}_{\mathrm{H}}^{\mathrm{NbH},\ nc}$, is given by 

\begin{equation}
\begin{multlined}
  \label{eq:Sbar_Hnc^NbHx}
\bar{S}_{\mathrm{H}}^{\mathrm{NbH},\ nc} = \frac{\partial S_{\mathrm{H}}^{\mathrm{NbH},\ nc}}{\partial \mathrm{x}} = k_B \sum_{i=1}^{3} \Biggr [ \frac{\hbar \nu_{\mathrm{H}}^i}{\mbox{\large\(%
k_BT(%
\)}e^{\frac{\hbar\nu_{\mathrm{H}^i}}{k_BT}}-1\mbox{\large\(%
)%
\)}} \\
- \mathrm{\log} \Big [1-e^{\frac{-\hbar\nu_{\mathrm{H}^i}}{k_BT}} \Big ] \Biggr ]
\end{multlined}
\end{equation}

Combining Eq. \ref{eq:Sbar_H^NbHx-1},  Eq. \ref{eq:Sbar_Hc^NbHx-final}, and Eq. \ref{eq:Sbar_Hnc^NbHx}, we get 

\begin{equation}
\begin{multlined}
\label{eq:Sbar_H^NbHx-final}
\bar{S}_{\mathrm{H}}^{\mathrm{NbH}} = -k_B\mathrm{\log}[\frac{\mathrm{x}}{R-\mathrm{x}}] + k_B \sum_{i=1}^{3} \Biggr [ \frac{\hbar \nu_{\mathrm{H}}^i}{\mbox{\large\(%
k_BT(%
\)}e^{\frac{\hbar\nu_{\mathrm{H}^i}}{k_BT}}-1\mbox{\large\(%
)%
\)}} \\
- \mathrm{\log}\Big[1-e^{\frac{-\hbar\nu_{\mathrm{H}^i}}{k_BT}} \Big] \Biggr]
\end{multlined}
\end{equation}

$\mu_H^{\mathrm{NbH}}$ in Eq. \ref{eq:mu_H^NbH-2} can be calculated by combining Eq. \ref{eq:Hbar_H^NbHx-final}, and Eq. \ref{eq:Sbar_H^NbHx-final} as

\begin{equation}
\begin{multlined}
\label{eq:mu_H^NbH-3}
\mu_H^{\mathrm{NbH}} = E_{\mathrm{H}} + \frac{1}{2} \sum_{i=1}^{3} \hbar \nu_H^i + 2\alpha\mathrm{x} + 3\beta\mathrm{x}^2 + 4\gamma\mathrm{x}^3 \\
+ k_BT\mathrm{log} \biggr[ \frac{\mathrm{x}}{R-\mathrm{x}}\biggr] + k_BT\mathrm{\log} \Big[1-e^{\frac{-\hbar \nu_{\mathrm{H}}^i}{k_BT}}\Big]
\end{multlined}
\end{equation}

\begin{figure}[h]
\centering
\includegraphics[width=\columnwidth]{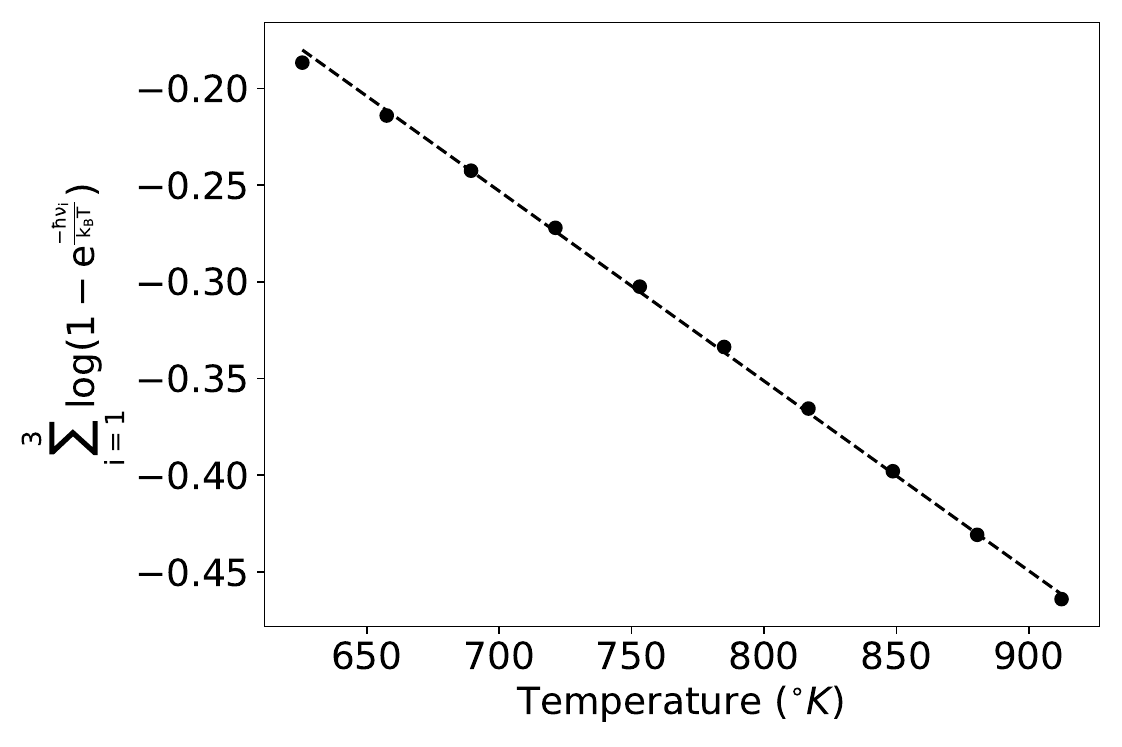}
\caption{Linear variation of the $\log [1-e^{\frac{-\hbar \nu_{\mathrm{H}}^i}{k_BT}}]$ term in Eq. \ref{eq:mu_H^NbH-3} for 625.65 K $<$ T $<$ 944.15 K}
\label{fig:4.7}       
\end{figure}

Next, we examine how $\sum_{i=1}^{3} \log \Big[1-e^{\frac{-\hbar \nu_{\mathrm{H}}^i}{k_BT}}\Big]$ varies with temperature in the range 625.65 K -- 944.15 K. We use the vibrational frequencies of an interstitial hydrogen atom in niobium calculated by Lee et al. \cite{leeUnderstandingDeviationsHydrogen2015a} to calculate $\sum_{i=1}^{3} \log \Big[1-e^{\frac{-\hbar \nu_{\mathrm{H}}^i}{k_BT}}\Big]$ and plot its variation in Fig. \ref{fig:4.7}. We observe that this function has a linear trend in this temperature range and perform OLS linear regression to find a linear fit. This fit is plotted in Fig. \ref{fig:4.7}. The linear relationship is given by Eq. \ref{eq:log_term_linear_fit}. \\

\begin{equation}
\begin{aligned}
\label{eq:log_term_linear_fit}
\sum_{i=1}^{3} \mathrm{log} \Big[1-e^{\frac{-\hbar \nu_{\mathrm{H}}^i}{k_BT}}\Big] = c + mT \\
\end{aligned}
\end{equation}

In Eq. \ref{eq:log_term_linear_fit}, m and c are fitting constants. For 625.65 K < T < 944.15 K, the values $m = -9.82 \times 10^{-4}$ and $c = 0.43$, correspond to the best linear fit. \\

Combining Eq. \ref{eq:mu_H^NbH-3} and Eq. \ref{eq:log_term_linear_fit}, we get 

\begin{equation}
\begin{multlined}
\label{eq:mu_H^NbH-final}
\mu_H^{\mathrm{NbH}} = E_{\mathrm{H}} + \frac{1}{2} \sum_{i=1}^{3} \hbar \nu_H^i + 2\alpha\mathrm{x} + 3\beta\mathrm{x}^2 + 4\gamma\mathrm{x}^3 \\
+ k_BT\mathrm{log} \biggr[\frac{\mathrm{x}}{R-\mathrm{x}}\biggr] + k_BT(c+ mT)
\end{multlined}
\end{equation}

We now turn our attention to calculating $\mu_{\mathrm{H}}^{\mathrm{H_2}}$, the chemical potential of hydrogen in the gas phase. As we did for the metal phase, we consider $H_{\mathrm{H_2}}^{\circ}$, the enthalpy of hydrogen in the gas phase and $S_{\mathrm{H_2}}^{\circ}$, the entropy of hydrogen in the gas phase as functions of temperature. $\mu_{\mathrm{H}}^{\mathrm{H_2}}$ can be expressed as

\begin{equation}
\begin{multlined}
\label{eq:mu_H^H2-2}
\mu_{\mathrm{H}}^{\mathrm{H_2}} = H_{\mathrm{H_2}}^{\circ}(T) - TS_{\mathrm{H_2}}^{\circ}(T) + k_BT\mathrm{log}(\frac{P}{P_0})
\end{multlined}
\end{equation}

\begin{figure}[h]
\includegraphics[width=\columnwidth]{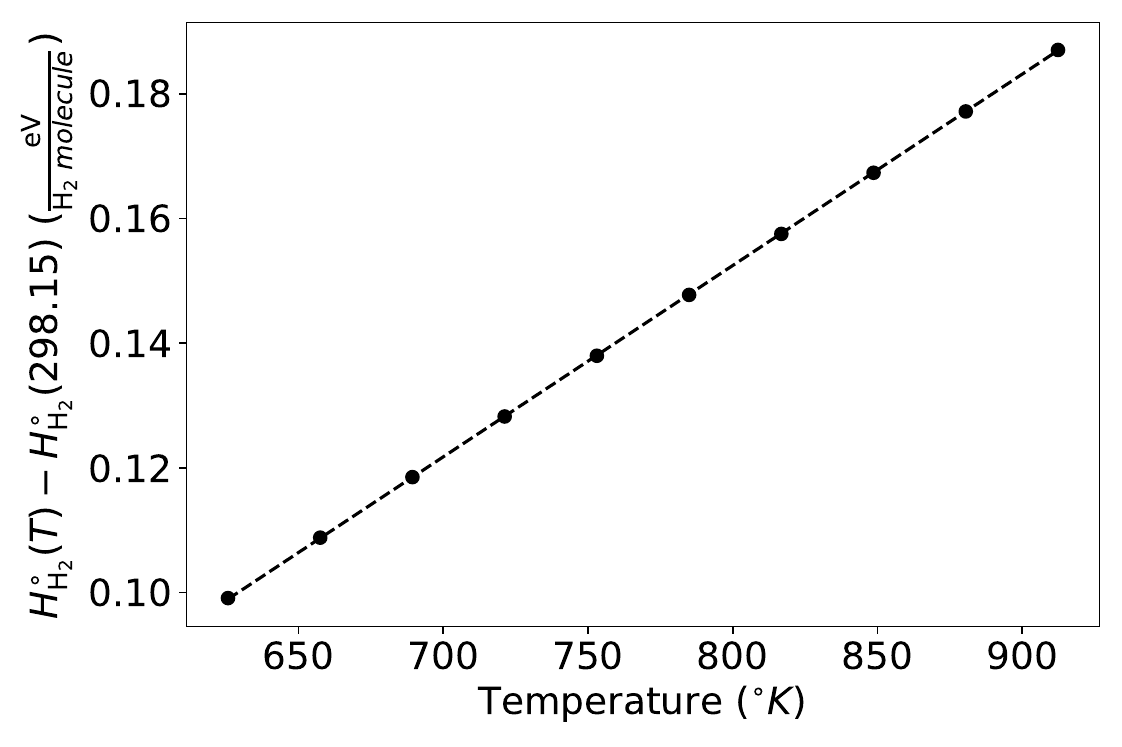}
\caption{A comparison of the $H_{\mathrm{H_2}}^{\circ}(T)\ - \ H_{\mathrm{H_2}}^{\circ}(298.15)$ values calculated using the exact functional form reported by NIST \citep{chase1998data} and a linear approximation for 625.65 K $<$ T $<$ 944.15 K}
\label{fig:4.8}       
\end{figure}

The functional form for the variation of $H_{\mathrm{H_2}}^{\circ}(T)$ with temperature has been reported by NIST \citep{chase1998data} referenced as $H_{\mathrm{H_2}}^{\circ}(T) - H_{\mathrm{H_2}}^{\circ}(298.15)$. When calculating the $H_{\mathrm{H_2}}^{\circ}(T) - H_{\mathrm{H_2}}^{\circ}(298.15)$ values using the functional form reported by NIST, we found that these values approximately have a linear dependence on temperature in the temperature range of interest (625.65 -- 944.15 K). Therefore, we perform an OLS-based linear regression to find a linear fit to the values calculated using the exact functional form reported by NIST. Fig. \ref{fig:4.8} shows that this linear fit is reasonable. \\

The linear fit for $H_{\mathrm{H_2}}^{\circ}(T) - H_{\mathrm{H_2}}^{\circ}(298.15)$ shown in Fig. \ref{fig:4.8} takes the form 

\begin{equation}
\begin{multlined}
  \label{eq:H_H2(T)}
H_{\mathrm{H_2}}^{\circ}(T) - H_{\mathrm{H_2}}^{\circ}(298.15) = H_{\mathrm{H_2}}^{\circ} + \Delta H_{\mathrm{H_2}}^{\circ}T \\ 
\text{for}\ 625.65 K < T < 944.15 K\\
\end{multlined}
\end{equation}

In Eq. \ref{eq:H_H2(T)}, $H_{\mathrm{H_2}}^{\circ}$ and $\Delta H_{\mathrm{H_2}}^{\circ}$ are fitting constants. For 625.65 K < T < 944.15 K, the values $H_{\mathrm{H_2}}^{\circ} = -0.093\ \frac{\mathrm{eV}}{\mathrm{H_2}\ molecule}$ and $\Delta H_{\mathrm{H_2}}^{\circ}\ =\ 3.07 \times 10^{-4}\ \frac{\mathrm{eV}}{K\ \mathrm{H_2}\ molecule}$, correspond to the best linear fit. \\

\begin{figure}[h]
\includegraphics[width=\columnwidth]{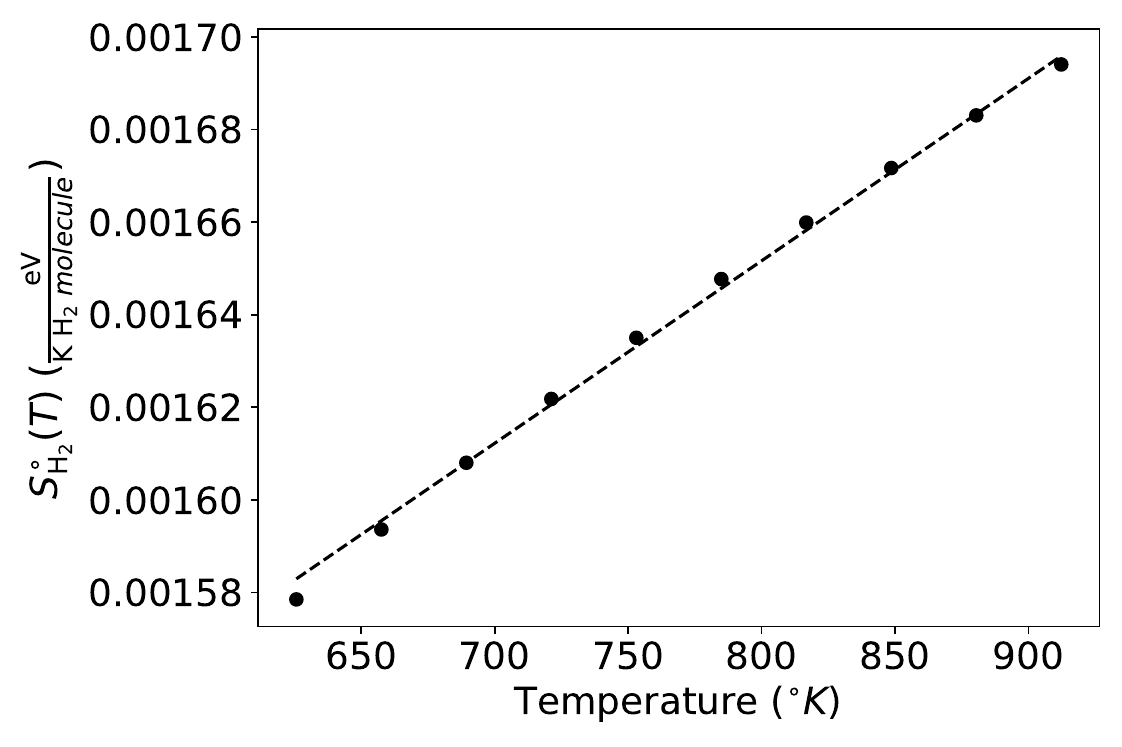}
\caption{A comparison of the $S_{\mathrm{H_2}}^{\circ}(T)$ values calculated using the exact functional form reported by NIST \citep{chase1998data} and a linear approximation for 625.65 K $<$ T $<$ 944.15 K}
\label{fig:4.9}       
\end{figure}

The functional form for the variation of $S_{\mathrm{H_2}}^{\circ}(T)$ with temperature has also been reported by NIST \citep{chase1998data}. We found that the values for $S_{\mathrm{H_2}}^{\circ}(T)$ calculated using this functional form approximately have a linear dependence on temperature for temperatures in the range 625.65 -- 944.15 K. Therefore, we performed an OLS-based linear regression to find a linear fit to the $S_{\mathrm{H_2}}^{\circ}(T)$ values. Fig. \ref{fig:4.9} shows the goodness of the linear approximation. \\

The linear fit for $S_{\mathrm{H_2}}^{\circ}(T)$ shown in Fig. \ref{fig:4.9} takes the form

\begin{equation}
\begin{aligned}
  \label{eq:S_H2(T)}
S_{\mathrm{H_2}}^{\circ}(T) = S_{\mathrm{H_2}}^{\circ} + \Delta S_{\mathrm{H_2}}^{\circ}T \\
\end{aligned}
\end{equation}

In Eq. \ref{eq:S_H2(T)}, $S_{\mathrm{H_2}}^{\circ}$ and $\Delta S_{\mathrm{H_2}}^{\circ}$ are fitting constants. For 625.65 K < T < 944.15 K, the values $S_{\mathrm{H_2}}^{\circ} =\ 1.34 \times 10^{-3} \ \frac{\mathrm{eV}}{K\ \mathrm{H_2}\ molecule}$ and $\Delta S_{\mathrm{H_2}}^{\circ}\ =\ 3.94 \times 10^{-7}\ \frac{\mathrm{eV}}{K^2\ \mathrm{H_2}\ molecule}$, correspond to the best linear fit. \\

Combining Eq. \ref{eq:mu_H^H2-2}, Eq. \ref{eq:H_H2(T)}, and  Eq. \ref{eq:S_H2(T)} we get 

\begin{equation}
\begin{multlined}
\label{eq:mu_H^H2-final}
\mu_{\mathrm{H}}^{\mathrm{H_2}} =  H_{\mathrm{H_2}}^{\circ} + H_{\mathrm{H_2}}^{\circ}(298.15) + (\Delta H_{\mathrm{H_2}}^{\circ} - S_{\mathrm{H_2}}^{\circ})T - \Delta S_{\mathrm{H_2}}^{\circ}T^2 \\
+ k_BT\mathrm{log}(\frac{P}{P_0})
\end{multlined}
\end{equation}

Using the expressions for $\mu_H^{\mathrm{NbH}}$ (Eq. \ref{eq:mu_H^NbH-final}) and $\mu_{\mathrm{H}}^{\mathrm{H_2}}$ (Eq. \ref{eq:mu_H^H2-final}) to satisfy the condition for chemical equilibrium in Eq. \ref{eq:chemeqbm}, we get 

\begin{equation}
\begin{multlined}
\label{eq:logP-relation}
\mathrm{log}(P^{\frac{1}{2}}) = \mathrm{log}(P_0^{\frac{1}{2}}) + \frac{1}{k_BT}[E_{\mathrm{H}} + \frac{1}{2} \sum_{i=1}^{3} \hbar \nu_H^i 
+ 2\alpha\mathrm{x} + 3\beta\mathrm{x}^2 + 4\gamma\mathrm{x}^3 \\
-\frac{1}{2}(H_{\mathrm{H_2}}^{\circ} + H_{\mathrm{H_2}}^{\circ}(298.15))] + \mathrm{log}\biggr[\frac{\mathrm{x}}{R-\mathrm{x}}\biggr] \\
+ (c + \frac{1}{2k_B}(S_{\mathrm{H_2}}^{\circ}-\Delta H_{\mathrm{H_2}}^{\circ})) 
+ (\frac{1}{2k_B}\Delta S_{\mathrm{H_2}}^{\circ} + m)T \\
\end{multlined}
\end{equation}

Eq.\ \ref{eq:logP-relation} can be simplified to

\begin{equation}
\begin{multlined}
\label{eq:P-relation-final}
P^{\frac{1}{2}} = P_0^{\frac{1}{2}} 
e^{\eta + \mu T + \frac{1}{k_BT}[H + 2\alpha\mathrm{x} + 3\beta\mathrm{x}^2 + 4\gamma\mathrm{x}^3] + \mathrm{log}[\frac{\mathrm{x}}{R-\mathrm{x}}]}
\\
\mathrm{Where}, \\
\eta = \frac{1}{2k_B}(S_{\mathrm{H_2}}^{\circ} -\Delta H_{\mathrm{H_2}}^{\circ}) + c \\
\mu = \frac{\Delta S_{\mathrm{H_2}}^{\circ}}{2k_B} + m \\
H = E_{\mathrm{H}} + \frac{1}{2} \sum_{i=1}^{3} \hbar \nu_H^i -\frac{1}{2}(H_{\mathrm{H_2}}^{\circ} + H_{\mathrm{H_2}}^{\circ}(298.15)) \\
R = R_0 + R_1T + R_2T^2 \\
\end{multlined}
\end{equation}

We now want to derive the expression for the relative partial molal enthalpy, $\bar{H}_{\mathrm{H}}^{\mathrm{NbH}} - \frac{1}{2} H_{\mathrm{H_2}}^{\circ}$, according to this model. Using Eq. \ref{eq:Hbar_H^NbHx-final} and Eq. \ref{eq:H_H2(T)}, we get

\begin{equation}
\begin{multlined}
\label{eq:H_H^NbH-model-derived}
\bar{H}_{\mathrm{H}}^{\mathrm{NbH}} - \frac{1}{2} H_{\mathrm{H_2}}^{\circ}(T) = E_{\mathrm{H}} + \frac{1}{2} \sum_{i=1}^{3} \hbar \nu_H^i +  \sum_{i=1}^{3} \frac{\hbar \nu_{\mathrm{H}}^i}{e^{\frac{\hbar \nu_{\mathrm{H}}^i}{k_BT}}-1} \\
+ 2\alpha\mathrm{x} + 3\beta\mathrm{x}^2 + 4\gamma\mathrm{x}^3 \\
- \frac{1}{2}(H_{\mathrm{H_2}}^{\circ} + H_{\mathrm{H_2}}^{\circ}(298.15) + \Delta H_{\mathrm{H_2}}^{\circ}T)
\end{multlined}
\end{equation}

The relative partial molal entropy, $\bar{S}_{\mathrm{H}}^{\mathrm{NbH}} - \frac{1}{2} S_{\mathrm{H_2}}^{\circ}(T)$, can be obtained from Eq. \ref{eq:Sbar_H^NbHx-final} and Eq. \ref{eq:S_H2(T)} 

\begin{equation}
\begin{multlined}
\label{eq:S_H^NbH-model-derived}
\bar{S}_{\mathrm{H}}^{\mathrm{NbH}} - \frac{1}{2} S_{\mathrm{H_2}}^{\circ}(T) = -k_B\mathrm{\log}\biggr[\frac{\mathrm{x}}{R-\mathrm{x}}\biggr] \\
+ k_B \sum_{i=1}^{3} \Biggr [ \frac{\hbar \nu_{\mathrm{H}}^i}{e^{\frac{\hbar \nu_{\mathrm{H}}^i}{k_BT}}-1} - \mathrm{\log}\Big[1-e^{\frac{-\hbar \nu_{\mathrm{H}}^i}{k_BT}}\Big] \Biggr] \\
- \frac{1}{2}(S_{\mathrm{H_2}}^{\circ} + \Delta S_{\mathrm{H_2}}^{\circ}T) \\
\end{multlined}
\end{equation} 

\section{Model Fitting}\label{sec4}

We now want to check if the derived expressions for relative partial molal enthalpy (Eq. \ref{eq:H_H^NbH-model-derived}) and relative partial molal entropy (Eq. \ref{eq:S_H^NbH-model-derived}) make a good fit to the corresponding values reported by Veleckis et al.\ \cite{veleckisThermodynamicPropertiesSystems1969a} Before we do that, we acknowledge that the values for these quantities reported by Veleckis et al.\ \cite{veleckisThermodynamicPropertiesSystems1969a}  were derived assuming that they do not depend on temperature. Without temperature dependence, the expression for the relative partial molal enthalpy in Eq. \ref{eq:H_H^NbH-model-derived} reduces to 

\begin{equation}
\begin{multlined}
\label{eq:H_H^NbH-model-derived-temp-independent}
\bar{H}_{\mathrm{H}}^{\mathrm{NbH}} - \frac{1}{2} H_{\mathrm{H_2}}^{\circ} = H + 2\alpha\mathrm{x} + 3\beta\mathrm{x}^2 + 4\gamma\mathrm{x}^3\ \\ 
Where,\ H = E_{\mathrm{H}} + \frac{1}{2} \sum_{i=1}^{3} \hbar \nu_H^i -\frac{1}{2}(H_{\mathrm{H_2}}^{\circ} + H_{\mathrm{H_2}}^{\circ}(298.15)) \\
\end{multlined}
\end{equation}

The expression for the relative partial molal entropy in Eq. \ref{eq:S_H^NbH-model-derived} reduces to 

\begin{equation}
\begin{multlined}
\label{eq:S_H^NbH-model-derived-temp-independent}
\bar{S}_{\mathrm{H}}^{\mathrm{NbH}} - \frac{1}{2} S_{\mathrm{H_2}}^{\circ} =  - k_B\mathrm{\log}\biggr[\frac{\mathrm{x}}{R_{mean}-\mathrm{x}}\biggr] - \frac{1}{2} S_{\mathrm{H_2}}^{\circ}
\end{multlined}
\end{equation}

In Eq.\ \ref{eq:H_H^NbH-model-derived-temp-independent}, $H_{\mathrm{H_2}}^{\circ} + H_{\mathrm{H_2}}^{\circ}(298.15)$ is a temperature-independent measure of the enthalpy of $\mathrm{H_2}$ gas in the experimental temperature range. And likewise in Eq. \ref{eq:S_H^NbH-model-derived-temp-independent}, $S_{\mathrm{H_2}}^{\circ}$ is a temperature-independent measure of the entropy of $\mathrm{H_2}$ gas. Additonally, in Eq. \ref{eq:S_H^NbH-model-derived-temp-independent}, $R_{mean}$ is a temperature-independent mean value of the number of interstitial sites per metal atom that is available for occupation in the experimental temperature range. \\

\begin{figure}[h]
\includegraphics[width=\columnwidth]{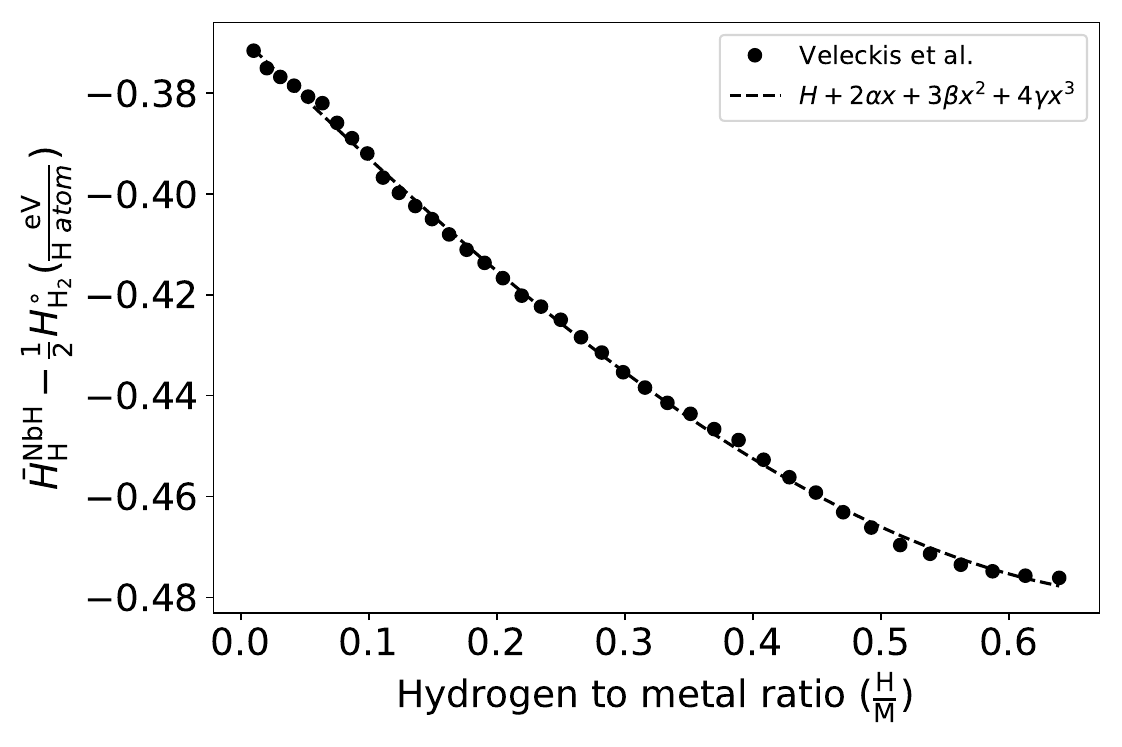}
\caption{Fitting Veleckis et al.'s \cite{veleckisThermodynamicPropertiesSystems1969a} relative partial molal enthalpy values to the expression for $\bar{H}_{\mathrm{H}}^{\mathrm{NbH}} - \frac{1}{2} H_{\mathrm{H_2}}^{\circ}$ in Eq. \ref{eq:H_H^NbH-model-derived-temp-independent} derived through the phenomenological model}
\label{fig:4.10}       
\end{figure}

Using an OLS treatment, we fit the relative partial molal enthalpy values reported by Veleckis et al.\ \cite{veleckisThermodynamicPropertiesSystems1969a} to the functional form for $\bar{H}_{\mathrm{H}}^{\mathrm{NbH}} - \frac{1}{2} H_{\mathrm{H_2}}^{\circ}$ given in Eq. \ref{eq:H_H^NbH-model-derived-temp-independent}. In Fig. \ref{fig:4.10}, we show that the fourth order dependence on $\mathrm{x}$ is an excellent fit to the original data. In Eq. \ref{eq:H_H^NbH-model-derived-values-temp-independent}, we report the values for the fitting constants corresponding to the best fit to the relative partial molal enthalpy data. 

\begin{equation}
\begin{multlined}
\label{eq:H_H^NbH-model-derived-values-temp-independent}
H = -0.37\ \mathrm{eV}/\mathrm{H}\ atom \\
\alpha = -0.12\ \frac{\mathrm{eV}/\mathrm{H}\ atom}{\mathrm{(H/Nb)}} \\
\beta = 0.02\ \frac{\mathrm{eV}/\mathrm{H}\ atom}{\mathrm{(H/Nb)}^2} \\
\gamma = 0.03\ \frac{\mathrm{eV}/\mathrm{H}\ atom}{\mathrm{(H/Nb)}^3} \\
\end{multlined}
\end{equation}

\begin{figure}[h]
\includegraphics[width=\columnwidth]{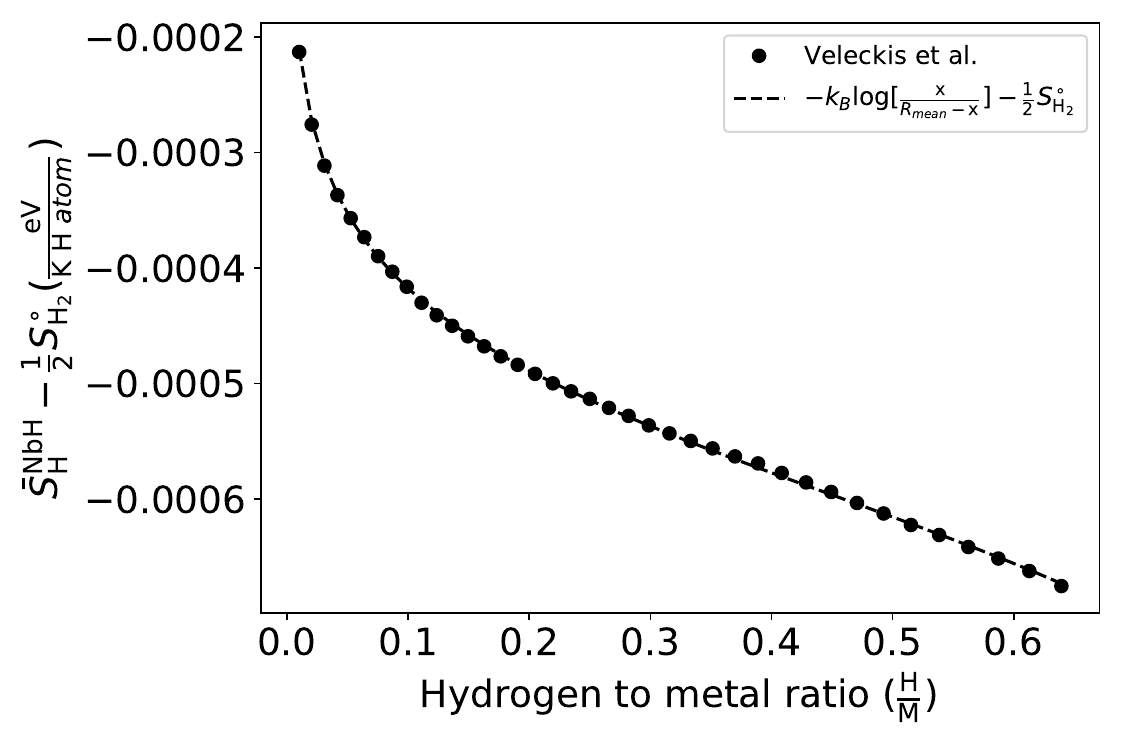}
\caption{Fitting Veleckis et al.'s \cite{veleckisThermodynamicPropertiesSystems1969a} relative partial molal entropy values to the expression for $\bar{S}_{\mathrm{H}}^{\mathrm{NbH}} - \frac{1}{2} S_{\mathrm{H_2}}^{\circ}$ in Eq. \ref{eq:S_H^NbH-model-derived-temp-independent} derived through the phenomenological model}
\label{fig:4.11}       
\end{figure}

Using an OLS treatment, we also fit the relative partial molal entropy values reported by Veleckis et al.\ \cite{veleckisThermodynamicPropertiesSystems1969a}  to the functional form for $\bar{S}_{\mathrm{H}}^{\mathrm{NbH}} - \frac{1}{2} S_{\mathrm{H_2}}^{\circ}$ given in Eq. \ref{eq:S_H^NbH-model-derived-temp-independent}. Fig. \ref{fig:4.11}, shows that this functional form makes an excellent fit to the entropy data reported by Veleckis et al.\ \cite{veleckisThermodynamicPropertiesSystems1969a} In Eq. \ref{eq:S_H^NbH-model-derived-values-temp-independent}, we report the values for the fitting constants corresponding to the best fit to the relative partial molal entropy data.

\begin{equation}
\begin{multlined}
\label{eq:S_H^NbH-model-derived-values-temp-independent}
S_{\mathrm{H_2}}^{\circ} = 1.2 \times 10^{-3} \ \frac{\mathrm{eV}}{K\ \mathrm{H_2}\ molecule} \\
R_{mean} = 0.90 \\
\end{multlined}
\end{equation}

We note that the $S_{\mathrm{H_2}}^{\circ}$ value derived by this analysis is in the order of the value ($1.34 \times 10^{-3} \ \frac{\mathrm{eV}}{K\ \mathrm{H_2}\ molecule}$) obtained by performing linear regression on the NIST reported values for $S_{\mathrm{H_2}}^{\circ}(T)$. This agreement provides an independent check on the validity of this methodology. \\

Without accounting for temperature dependencies, we have derived expressions for $\bar{H}_{\mathrm{H}}^{\mathrm{NbH}} - \frac{1}{2} H_{\mathrm{H_2}}^{\circ}$, and $\bar{S}_{\mathrm{H}}^{\mathrm{NbH}} - \frac{1}{2} S_{\mathrm{H_2}}^{\circ}$ given by equations \ref{eq:H_H^NbH-model-derived-temp-independent} and \ref{eq:S_H^NbH-model-derived-temp-independent} respectively. Combining these expressions with equations \ref{eq:veleckis_eqbm_condition}, \ref{eq:relative_enthalpy_B_relation}, \ref{eq:relative_entropy_A_relation}, we derive an expression for the equilibrium pressure given by Eq. \ref{eq:P-relation-temp-independent}.

\begin{equation}
\begin{multlined}
\label{eq:P-relation-temp-independent}
P^{\frac{1}{2}} = P_0^{\frac{1}{2}}\huge{e}^{\biggr[\frac{1}{k_BT}[H + 2\alpha\mathrm{x} + 3\beta\mathrm{x}^2 + 4\gamma\mathrm{x}^3] + \frac{S_{\mathrm{H_2}}^{\circ}}{2k_B} + \mathrm{log}\big[\frac{\mathrm{x}}{R_{mean}-\mathrm{x}}\big]\biggr]} \\
\end{multlined}
\end{equation}

We note that in the absence of temperature dependencies of the enthalpy and entropy of hydrogen in the metal phase and gas phase, the expression for equilibrium pressure in Eq. \ref{eq:P-relation-final} reduces to the one in Eq. \ref{eq:P-relation-temp-independent}. This reduction confirms the expected equivalence in the two approaches to derive the equilibrium pressure in the limit where the relative partial molal enthalpy and the relative partial molal entropy are temperature independent. \\

\begin{figure}[h]
\includegraphics[width=\columnwidth]{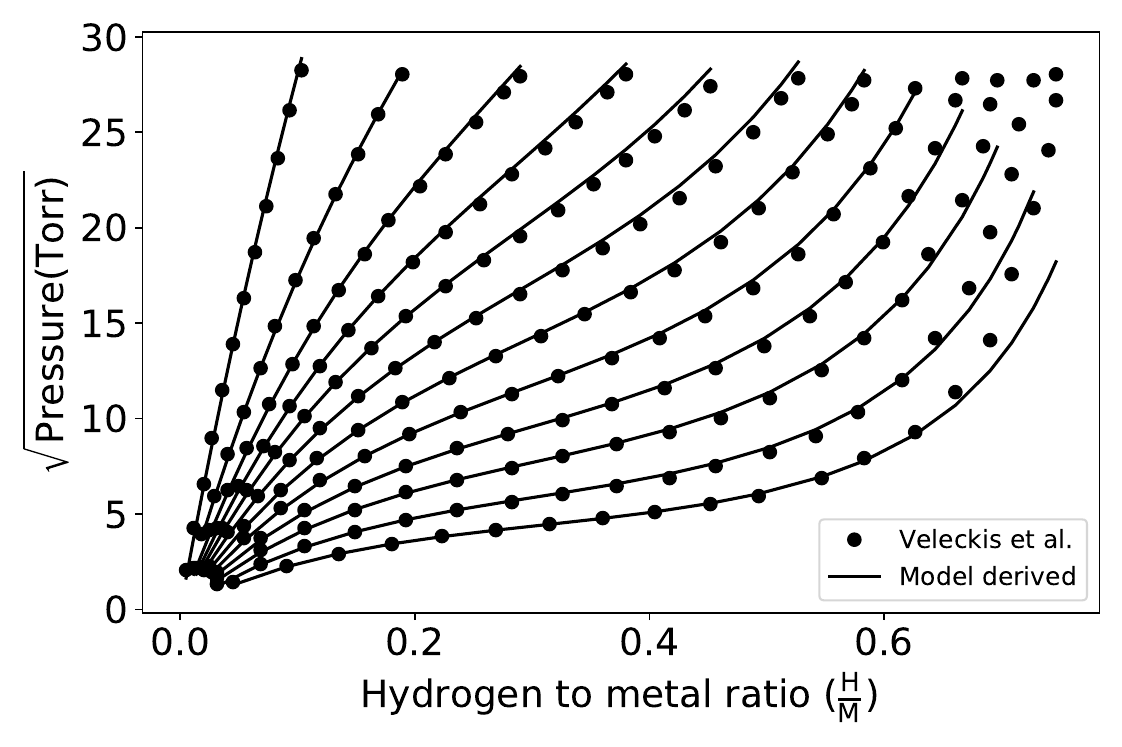}
\caption{A comparison of absorption isotherms reported by Veleckis et al.\ \cite{veleckisThermodynamicPropertiesSystems1969a} and those calculated using Eq. \ref{eq:P-relation-temp-independent} by substituting the fitting parameters values reported in Eq. \ref{eq:H_H^NbH-model-derived-values-temp-independent} and Eq. \ref{eq:S_H^NbH-model-derived-values-temp-independent}}
\label{fig:4.12}       
\end{figure}

Using the values for the fitting constants $H$, $\alpha$, $\beta$, $\gamma$, $S_{\mathrm{H_2}}^{\circ}$, and $R_{mean}$ in Eq. \ref{eq:P-relation-temp-independent}, we can calculate the equilibrium pressures for varying x and $T$. In Fig. \ref{fig:4.12}, we plot the calculated pressures compared to Veleckis et al.'s \cite{veleckisThermodynamicPropertiesSystems1969a} original data. \\

From Fig. \ref{fig:4.12}, it is evident that the pressures calculated are in much better agreement with the original data for 0 $<$ $\mathrm{x}$ $<$ 0.64, as this corresponds to the range in which the fitting for the $\bar{H}_{\mathrm{H}}^{\mathrm{NbH}} - \frac{1}{2} H_{\mathrm{H_2}}^{\circ}$ and $\bar{S}_{\mathrm{H}}^{\mathrm{NbH}} - \frac{1}{2} S_{\mathrm{H_2}}^{\circ}$ values was performed. For $\mathrm{x}$ $>$ 0.64, the deviation of the calculated pressures from the original data is quite drastic, as they are not represented in the fit. \\

We introduce an error function, $\lambda$, given by Eq. \ref{eq:loss-function} to quantify the deviation of the model-predicted isotherms from the experimentally reported absorption isotherms. \\


\begin{equation}
\label{eq:loss-function}
\lambda =\sum_{i=1}^{N} \Biggr[ \frac{1}{\sigma_{P^{\frac{1}{2}}}^2} (P^{\frac{1}{2}}(\mathrm{x_i}) - \boldsymbol{P_{i}^{\frac{1}{2}})}^2 + \frac{1}{\sigma_\mathrm{x_i^2}}(\mathrm{x_i}-\mathbf{x_i})^2 + \frac{1}{\sigma_T^2}(T_i-\boldsymbol{T_i})^2 \Biggr ]
\end{equation}


In Eq. \ref{eq:loss-function}, $\boldsymbol{P_i^{\frac{1}{2}}}$, $\mathbf{x_i}$, and $\boldsymbol{T_i}$ represent the experimentally observed variables and $P^{\frac{1}{2}}(\mathrm{x_i})$, $\mathrm{x_i}$, and $T_i$ represent the underlying values that lie on the isotherm. N is the total number of observations, which equals 207. We assume that the error in measuring $\boldsymbol{P^{\frac{1}{2}}_i}$, $\mathbf{x_i}$, and $\boldsymbol{T_i}$ follow a normal distribution with the standard deviations given by $\sigma_{P^{\frac{1}{2}}}$, $\sigma_{\mathrm{x}}$, and $\sigma_T$ respectively. We expect that the error made in digitizing the absorption isotherms from Veleckis et al.\ \cite{veleckisThermodynamicPropertiesSystems1969a} to be the biggest source of error for $\boldsymbol{P_i^{\frac{1}{2}}}$ and $\mathbf{x_i}$ and estimate $\sigma_{P^{\frac{1}{2}}}$ to be 0.07 and $\sigma_{\mathrm{x}}$ to be 0.0014. We estimate $\sigma_T$ to be 0.1 based on the precision of the measurements made by Veleckis et al. We will discuss a fitting procedure that minimizes $\lambda$ for the given set of observations from Veleckis et al.\ \cite{veleckisThermodynamicPropertiesSystems1969a} later in this work. However, our first objective is to use $\lambda$ to quantify the deviation of the model-predicted isotherms from the experimental absorption isotherms. \\

The root means square error (RMSE) is a metric that normalizes the error by the number of observations, N. It is related to $\lambda$ as

\begin{equation}
\label{eq:RMSE}
\mathrm{RMSE} = \sqrt{\frac{\lambda}{\mathrm{N}}} 
\end{equation}

Using the definition of $\lambda$ in Eq. \ref{eq:loss-function}, we calculate the RMSE values for each isotherm in Fig. \ref{fig:4.12}. The isotherms at lower temperatures have larger RMSE values as they have more data points with $\mathrm{x}$ $>$ 0.64. The isotherm corresponding to T = 625.65 K has the largest RMSE of 13.72. The isotherm corresponding to T = 858.95 K has the smallest $\lambda$ value of 0.77. The RMSE value calculated over all data points is 5.92. \\

The expression for $P^{\frac{1}{2}}$ in Eq. \ref{eq:P-relation-temp-independent} can be thought of as $P^{\frac{1}{2}} = F(\mathrm{x}, T)$. Therefore, instead of calculating fitting constants from fitting the relative partial molal enthalpy and relative partial mola entropy data, we can calculate them by directly fitting the equilibrium pressures from the isotherms reported by Veleckis et al.\ \cite{veleckisThermodynamicPropertiesSystems1969a} to the expression in Eq. \ref{eq:P-relation-temp-independent}. \\

We perform this fitting procedure using orthogonal distance regression (ODR) \cite{boggsOrthogonalDistanceRegression1990} as implemented in Scipy's ODR package for Python \cite{OrthogonalDistanceRegression2021}. We chose ODR as our fitting procedure because OLS only accounts for uncertainties in the response variable, which is the equilibrium pressure in this case. We use ODR to account for uncertainties in the equilibrium pressure, hydrogen concentration, and temperature measurements. The objective function to be minimized in the ODR fitting is the weighted sum of squares error, $\lambda$ given in Eq. \ref{eq:loss-function}. \\

\begin{table}[h]
\caption{A comparison of model parameters and RMSE Values obtained following the fitting procedure outlined by Veleckis et al.\ \cite{veleckisThermodynamicPropertiesSystems1969a} and those obtained by fitting to the phenomenological model without accounting for temperature dependence, as in Eq. \ref{eq:P-relation-temp-independent}.}
    \begin{tabular}{ccc} 
    \hline
    Fitting variables & $\bar{H}_{\mathrm{H}}^{\mathrm{NbH}} - \frac{1}{2} H_{\mathrm{H_2}}^{\circ}$ and $\bar{S}_{\mathrm{H}}^{\mathrm{NbH}} - \frac{1}{2} S_{\mathrm{H_2}}^{\circ}$ & $P^{\frac{1}{2}}, \mathrm{x}, T$ \\ 

    & & \\


    Fitting method & OLS & ODR \\ 

    \hline 

    & & \\

    H ($\frac{\mathrm{eV}}{\mathrm{H}\ atom}$)& -0.37 & -0.36 \\ 

    & & \\


    $\alpha\ (\frac{\mathrm{eV}}{\mathrm{(H/Nb)}^2})$  & -0.12 & -0.14  \\ 

    & & \\

    $\beta\ (\frac{\mathrm{eV}}{\mathrm{(H/Nb)}^3})$ & 0.02 & 0.02 \\ 

    & & \\

    $\gamma\ (\frac{\mathrm{eV}}{\mathrm{(H/Nb)}^4})$ & 0.03 & 0.01 \\ 

    & & \\

    $R_{mean}$ & 0.90 & 0.80 \\ 

    & & \\

    $S_{\mathrm{H_2}}^{\circ}$ ($\frac{\mathrm{eV}}{K\ \mathrm{H}\ atom}$) & $1.2 \times 10^{-3}$ & $1.1 \times 10^{-3}$ \\

    & & \\
     
    RMSE & 5.92 & 1.03 \\

    & & \\

    \hline
    \end{tabular}
 \label{tab:4.1}
\end{table}

In Table \ref{tab:4.1}, we list the model parameters H, $\alpha$, $\beta$, $\gamma$, $S_{\mathrm{H_2}}^{\circ}$, and $R_{mean}$ representative of the best fit to the data from Veleckis et al.\ \cite{veleckisThermodynamicPropertiesSystems1969a} For comparison, we list the values reported in Eq. \ref{eq:H_H^NbH-model-derived-values-temp-independent} and Eq. \ref{eq:S_H^NbH-model-derived-values-temp-independent} corresponding to the best fits to the relative partial molal enthalpy and relative partial molal entropy data reported by Veleckis et al.\ \cite{veleckisThermodynamicPropertiesSystems1969a}. We also list the RMSE values corresponding to the two fitting methods in Table \ref{tab:4.1}. \\

Table \ref{tab:4.1} shows that both fitting methods yield similar values for all the fitting constants. However, fitting the equilibrium pressures from the isotherms reported by Veleckis et al.\ \cite{veleckisThermodynamicPropertiesSystems1969a} to the expression $P^{\frac{1}{2}} = F(\mathrm{x}, T)$ in Eq. \ref{eq:P-relation-temp-independent} using ODR, results in a much lower RMSE value compared to using the fitting procedure outlined by Veleckis et al.\ \cite{veleckisThermodynamicPropertiesSystems1969a}. This reduction in the RMSE value is especially impressive considering that Eq. \ref{eq:P-relation-temp-independent} only involves six free parameters, whereas Veleckis et al.'s \cite{veleckisThermodynamicPropertiesSystems1969a} fitting method involved 80 free parameters.

\begin{figure}[h]
\includegraphics[width=\columnwidth]{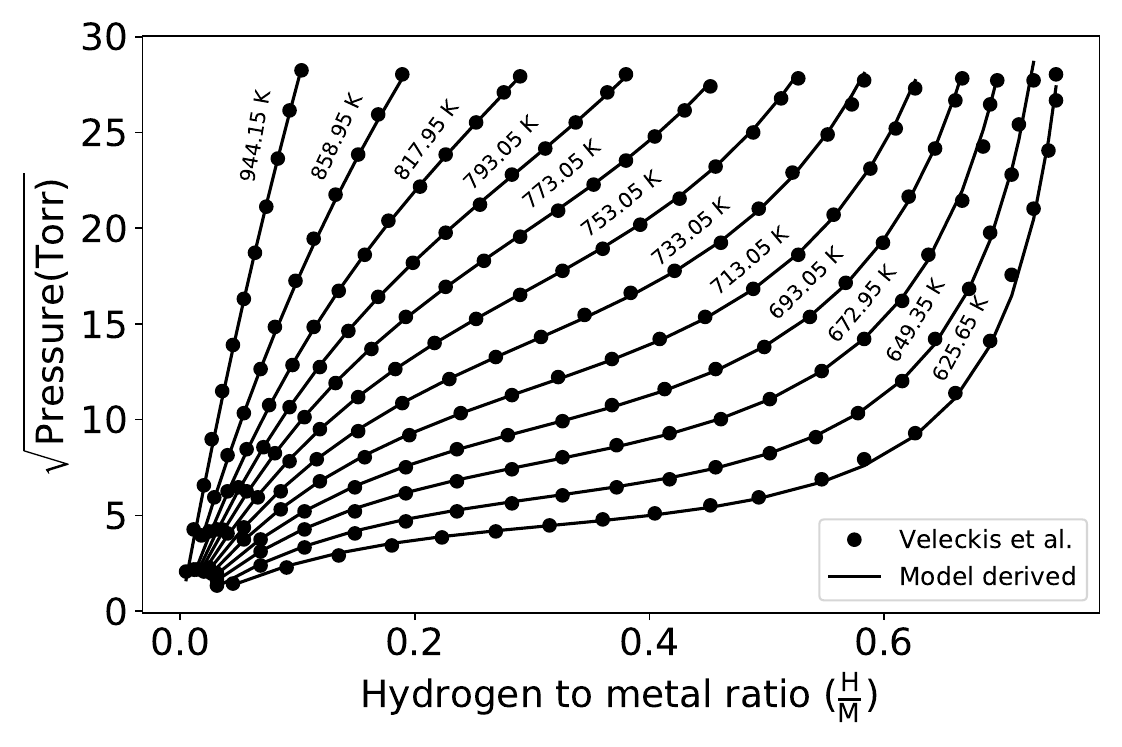}
\caption{A comparison of absorption isotherms reported by Veleckis et al.\ \cite{veleckisThermodynamicPropertiesSystems1969a} and those predicted by our model in the absence of remperature dependence of $\bar{H}_{\mathrm{H}}^{\mathrm{NbH}} - \frac{1}{2} H_{\mathrm{H_2}}^{\circ}$ and $\bar{S}_{\mathrm{H}}^{\mathrm{NbH}} - \frac{1}{2} S_{\mathrm{H_2}}^{\circ}$}
\label{fig:4.13}       
\end{figure}

In Fig. \ref{fig:4.13}, we plot the pressures calculated using Eq. \ref{eq:P-relation-temp-independent} with model parameters values in Tab. \ref{tab:4.1} in comparison with original data from Veleckis et al.\ \cite{veleckisThermodynamicPropertiesSystems1969a}. Comparing Fig. \ref{fig:4.13} and Fig. \ref{fig:4.12}, it is visibly obvious that fitting the experimental equilibrium pressure data to the expression in Eq. \ref{eq:P-relation-temp-independent} leads to a better fit. \\

Our next goal is to investigate if the phenomenological model developed here describes the absorption isotherms from Veleckis et al. \cite{veleckisThermodynamicPropertiesSystems1969a} when the temperature corrections to the enthalpy and entropy of hydrogen in the gas and metal phases are included. Therefore, we return our attention to the expression for the equilibrium pressure derived through the phenomenological model in Eq. \ref{eq:P-relation-final}. Eq. \ref{eq:P-relation-final} allows us to calculate the equilibrium pressure as $P^{\frac{1}{2}} = G(\mathrm{x}, T)$, and has up to nine free parameters ($\eta$, $\mu$, H, $\alpha$, $\beta$, $\gamma$, $R_0$, $R_1$, and $R_2$). Our next step is to fit the absorption isotherm data from Veleckis et al.\ \cite{veleckisThermodynamicPropertiesSystems1969a} to the expression for $P^{\frac{1}{2}}$ in Eq. \ref{eq:P-relation-final}. \\

We recall from Eq. \ref{eq:P-relation-final} that the terms $\eta$ and $\mu$ in Eq. \ref{eq:P-relation-final} are given by

\begin{equation}
\begin{multlined}
\label{eq:eta-and-mu}
\eta = \frac{1}{2k_B}(S_{\mathrm{H_2}}^{\circ} -\Delta H_{\mathrm{H_2}}^{\circ}) + c  \\
\mu = \frac{\Delta S_{\mathrm{H_2}}^{\circ}}{2k_B} + m \quad \quad \quad \ 
\end{multlined}
\end{equation}

The values for $S_{\mathrm{H_2}}^{\circ}$, $\Delta H_{\mathrm{H_2}}^{\circ}$, $\Delta S_{\mathrm{H_2}}$, c, and m were estimated earlier from the linear regression analysis shown in Figures \ref{fig:4.7}, \ref{fig:4.8}, and \ref{fig:4.9}. Using these values in Eq. \ref{eq:eta-and-mu}, we estimate $\boldsymbol{\eta} = 6.407$ and $\boldsymbol{\mu} = 1.3 \times 10^{-3}$. However, these estimates can have uncertainties, and the values for $\eta$ and $\mu$ in the best fit need not be equal to these estimates. To account for the difference in $\eta$ and $\mu$ values from the estimates $\boldsymbol{\eta}$ and $\boldsymbol{\mu}$, we modify the loss function in Eq. \ref{eq:loss-function} to include the weighted sum of squares of the differences as shown in Eq. \ref{eq:loss-function-A+BT}.

\begin{equation}
\begin{multlined}
\label{eq:loss-function-A+BT}
\lambda =\sum_{i=1}^{N} \Biggr[ \frac{1}{\sigma_{P^{\frac{1}{2}}}^2} (P^{\frac{1}{2}}(\mathrm{x_i}) - \boldsymbol{P_{i}^{\frac{1}{2}})}^2 + \frac{1}{\sigma_\mathrm{x_i^2}}(\mathrm{x_i}-\mathbf{x_i})^2 + \\
\frac{1}{\sigma_T^2}(T_i-\boldsymbol{T_i})^2 \Biggr] + \frac{1}{\sigma_{\eta}^2}(\eta - \boldsymbol{\eta})^2 + \frac{1}{\sigma_{\mu}^2}(\mu - \boldsymbol{\mu})^2
\end{multlined}
\end{equation}

Where $\sigma_{\eta}$ and $\sigma_{\mu}$ in Eq. \ref{eq:loss-function-A+BT} are the uncertainties in the estimates $\boldsymbol{\eta} = 6.407$ and $\boldsymbol{\mu} = 1.3 \times 10^{-3}$. The uncertainties in $\boldsymbol{\eta}$ and $\boldsymbol{\mu}$ can come from several sources. Firstly, the uncertainties in the $H_{\mathrm{H_2}}^{\circ}(T)$ and $S_{\mathrm{H_2}}^{\circ}(T)$ reported by NIST \citep{chase1998data}. Secondly the uncertainties from estimating $\Delta H_{\mathrm{H_2}}^{\circ}$, $S_{\mathrm{H_2}}^{\circ}$, and $\Delta S_{\mathrm{H_2}}^{\circ}$ using linear regression. Thirdly, the uncertainties in estimates for c and m and the estimation of the vibrational frequencies of an interstitial hydrogen atom in niobium by Lee et al. \cite{leeUnderstandingDeviationsHydrogen2015a}. Finally, assuming that all hydrogen atoms in the niobium-hydrogen system have independent vibrational modes and the vibrational frequencies of all hydrogen atoms are identical. \\

It would be instructive to minimize $\lambda$ in Eq. \ref{eq:loss-function-A+BT} for a spectrum of values $\sigma_{\eta}$ and $\sigma_{\mu}$ ranging from zero to infinity. However, in this work, we will limit our focus to the two extreme limits, $\sigma_{\eta} = \sigma_{\mu} = 0$ and $\sigma_{\eta} = \sigma_{\mu} = \infty$. \\

\begin{figure}[h]
\includegraphics[width=\columnwidth]{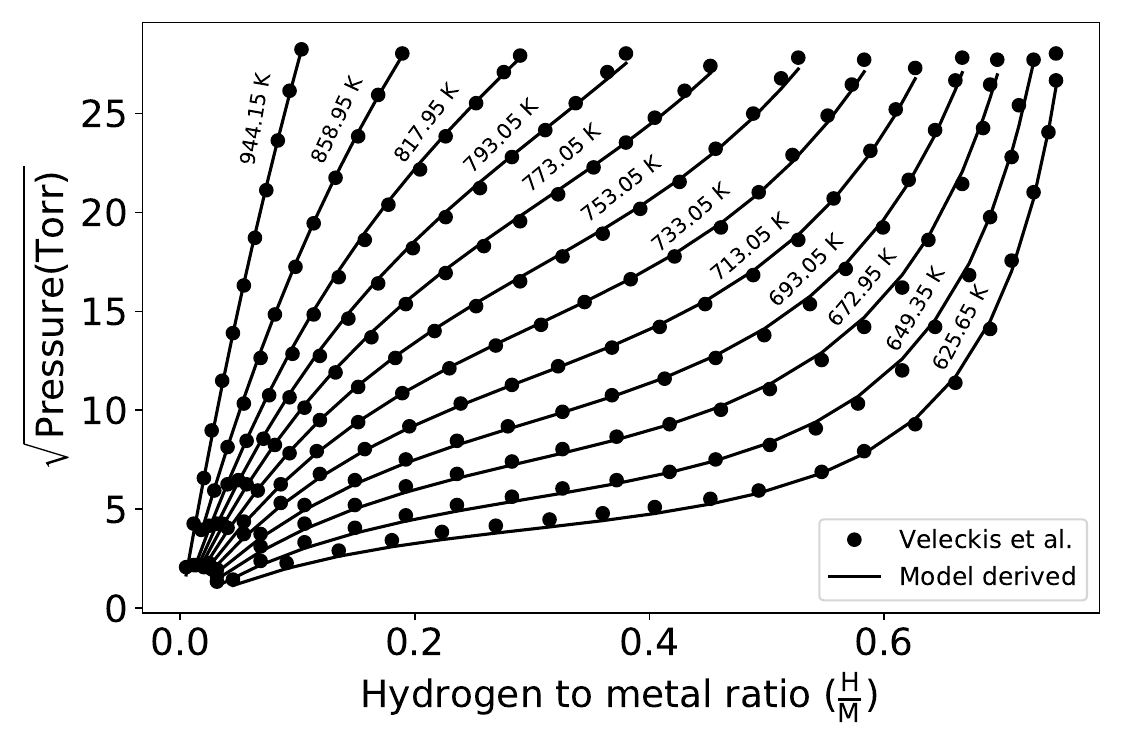}
\caption{A comparison of absorption isotherms reported by Veleckis et al.\ \cite{veleckisThermodynamicPropertiesSystems1969a} and those predicted by our model fixing $\eta$ and $\mu$ to 
the estimates $\boldsymbol{\eta}$ and $\boldsymbol{\mu}$ respectively}
\label{fig:4.14}       
\end{figure}

In the first limit, $\sigma_{\eta} = \sigma_{\mu} = 0$, i.e. there is no uncertainty in the estimates for $\boldsymbol{\eta}$ and $\boldsymbol{\mu}$. We implement this limit by fixing the values for $\eta$ and $\mu$ when fitting the expression for $P^{\frac{1}{2}}$ in Eq. \ref{eq:P-relation-final} to the experimental data. In this limit, the phenomenological model has 7 free parameters - H, $\alpha$, $\beta$, $\gamma$, $R_0$, $R_1$, $R_2$. We fit this model to the experimental data using ODR. In Fig. \ref{fig:4.14}, we present the ${P}^{\frac{1}{2}}$ reported by Veleckis et al.\ \cite{veleckisThermodynamicPropertiesSystems1969a} and the values corresponding to the model's best fit to the experimental data. Table \ref{tab:4.2} presents the model's parameters and the RMSE corresponding to the best fit.

\FloatBarrier
\FloatBarrier

\begin{figure}[h]
\includegraphics[width=\columnwidth]{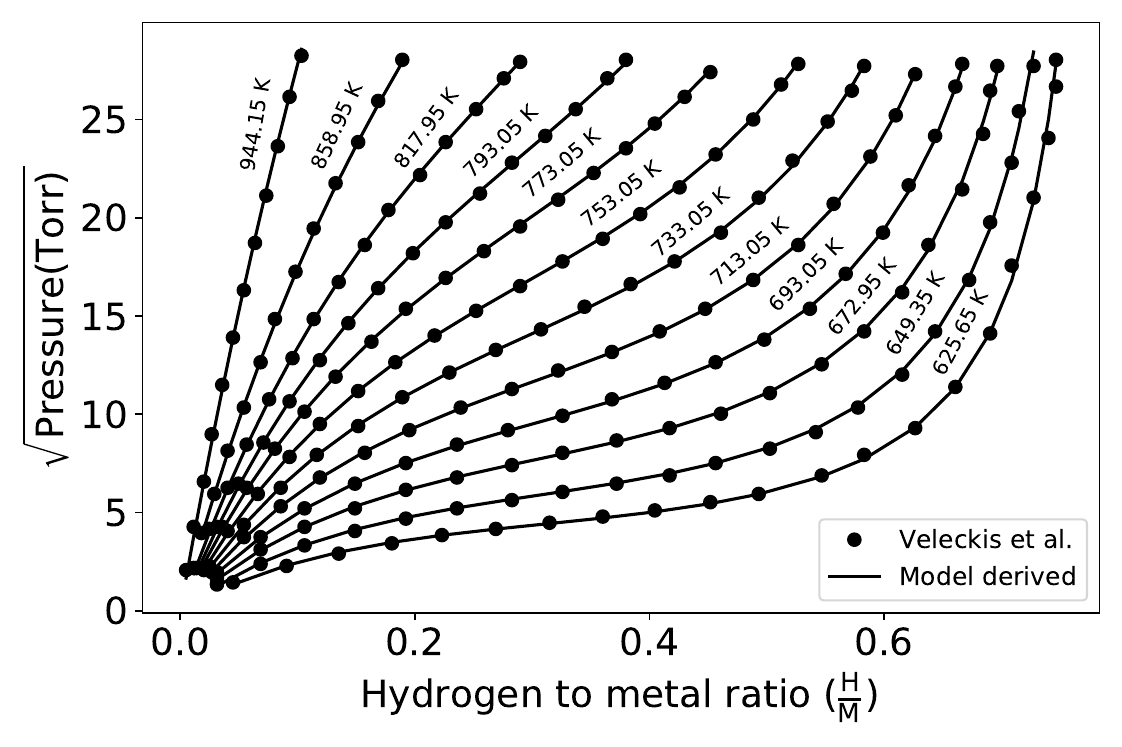}
\caption{A comparison of absorption isotherms reported by Veleckis et al.\ \cite{veleckisThermodynamicPropertiesSystems1969a} and those predicted by our model with $\eta$ and $\mu$ set as free parameters}
\label{fig:4.15}       
\end{figure}

In the second limit, $\sigma_{\eta} = \sigma_{\mu} = \infty$, i.e., there is a very large uncertainty in the estimates for $\boldsymbol{\eta}$ and $\boldsymbol{\mu}$. We implement this limit by setting $\eta$ and $\mu$ as free parameters when fitting the expression for $P^{\frac{1}{2}}$ in Eq. \ref{eq:P-relation-final} to the experimental data. In this limit, the phenomenological model has 9 free parameters - H. $\alpha$, $\beta$, $\gamma$, $R_0$, $R_1$, $R_2$, $\eta$, and $\mu$. Once again, we fit the model to the experimental data using ODR. In Fig. \ref{fig:4.15}, we present the ${P}^{\frac{1}{2}}$ from the experiments and the best fit to the experimental data. In Table \ref{tab:4.2}, we present the model's parameters and the RMSE corresponding to the best fit next to the values corresponding to best the fit when $\eta$ and $\mu$ were fixed.

\FloatBarrier
\FloatBarrier

From Fig. \ref{fig:4.15}, we note that the quality of the model's fit to the experimental data is better when we set $\eta$ and $\mu$ as free parameters compared to the fit shown in Fig. \ref{fig:4.14} where $\eta$ and $\mu$ were set to $\boldsymbol{\eta}$ and $\boldsymbol{\mu}$ respectively. The RMSE values listed in Table \ref{tab:4.2} for the two fits are consistent with this conclusion. A better fit is expected for the second limit of the model because of the extra degrees of freedom in $\eta$ and $\mu$. \\

Finally, we point out that the RMSE value for the second limit of the model (0.85) is lower in comparison to the RMSE value reported in the second column in Table \ref{tab:4.1} (1.03). This difference indicates the importance of including temperature corrections to the relative partial molal enthalpy and relative partial molal entropy in our phenomenological model.

\FloatBarrier
\begin{table}[h]
\caption{A comparison of model parameters and RMSE values obtained by fitting absorption isotherm data from Veleckis et al.\ \cite{veleckisThermodynamicPropertiesSystems1969a} to the phenomenological model under the two limits - fixing the $\eta$ and $\mu$ values to the estimates $\boldsymbol{\eta}$ and $\boldsymbol{\mu}$, and setting them as free parameters. Both models account for temperature dependence.}
\begin{tabular}{cccc} 
\hline
$\eta$ and $\mu$ & Fixed  & Free \\
\hline

& & \\

$H (\mathrm{eV}/\mathrm{H}\ atom)$ & -0.40 & -0.36 \\

& & \\

$\alpha\ (\frac{\mathrm{eV}/\mathrm{H}\ atom}{\mathrm{H/Nb}})$  & -0.12 & -0.14  \\

& & \\

$\beta\ (\frac{\mathrm{eV}/\mathrm{H}\ atom}{\mathrm{(H/Nb)}^2})$ & 0.02 & 0.03 \\ 

& & \\

$\gamma\ (\frac{\mathrm{eV}/\mathrm{H}\ atom}{\mathrm{(H/Nb)}^3})$ & 0.02 & 0.01 \\

& & \\

$R_0$ & 2.06 & 0.78 \\

& & \\

$R_1$ & -$4.3 \times 10^{-3}$ & -$1.8 \times 10^{-5}$ \\

& & \\

$R_2$ & $3.6 \times 10^{-6}$ & $9.8 \times 10^{-8}$ \\

& & \\

$R_{mean}$ & 0.99 & 0.82 \\

& & \\
 
$\eta$ & 6.41 & 6.69 \\ 

& & \\

$\mu$ & $1.3 \times 10^{-3}$ &  $1.3 \times 10^{-4}$ \\

& & \\

RMSE & 2.28 & 0.85 \\

& & \\ 

\hline \\
\end{tabular}
\label{tab:4.2}
\end{table}
\FloatBarrier

\section{Interpretation of Model Parameters}\label{sec4}

The model parameter values presented in Table \ref{tab:4.2} contain fundamental information regarding hydrogen absorption in niobium. Here, we will explore the information that can be derived from these parameter values. \\ 

From Eq. \ref{eq:Ch4-E_NbHx}, we recall that $\alpha$, $\beta$, and $\gamma$ are all related to the interactions between the hydrogen atoms. From Table \ref{tab:4.2}, we note that the $\alpha$ value is negative for both limits studied, and the values for $\beta$ and $\gamma$ are positive. From these observations, we infer that the interactions between the hydrogen atoms are a combination of attractive and repulsive components. Further, the non-zero values for $\beta$ and $\gamma$ indicate that the interactions are of many-body nature. These inferences are consistent with our previous work's conclusions about the interactions between interstitial hydrogen atoms in niobium based on density functional theory calculations \citep{ramachandranProbingInteractionsInterstitial2020}. \\

We can also derive the values for the relative partial molal enthalpy of hydrogen ($\Delta H_{\mathrm{x}}$) and the relative partial molal entropy of hydrogen ($\Delta S_{\mathrm{x}}$) from these model parameters. We recall that Veleckis et al.\ \cite{veleckisThermodynamicPropertiesSystems1969a} did not account for temperature dependencies when calculating $\Delta H_{\mathrm{x}}$ and $\Delta S_{\mathrm{x}}$. To make the values for $\Delta H_{\mathrm{x}}$ and $\Delta S_{\mathrm{x}}$ derived from this model comparable to the values reported by Veleckis et al. \cite{veleckisThermodynamicPropertiesSystems1969a}, here we will assume that $\bar{H}_{\mathrm{H}}^{\mathrm{NbH}}$, $H_{\mathrm{H_2}}^{\circ}$, $\bar{S}_{\mathrm{H}}^{\mathrm{NbH}}$, and $S_{\mathrm{H_2}}^{\circ}$ do not depend on temperature. \\

The relative partial molal enthalpy of hydrogen at a hydrogen-to-metal ratio of x is given by Eq. \ref{eq:Ch4-relative-partial-molal-enthalpy}.

\begin{equation}
\label{eq:Ch4-relative-partial-molal-enthalpy}
\Delta H_{\mathrm{x}} = \bar{H}_{\mathrm{H}}^{\mathrm{NbH}}(\mathrm{x}) - \frac{1}{2} H_{\mathrm{H_2}}^{\circ}
\end{equation}

In the absence of temperature dependence, we can use the expression for $\bar{H}_{\mathrm{H}}^{\mathrm{NbH}}(\mathrm{x})$ in Eq. \ref{eq:H_H^NbH-model-derived-temp-independent} to estimate $\Delta H_{\mathrm{x}}$ as

\begin{equation}
\label{eq:relative-partial-molal-enthalpy-temp-independent}
\Delta H_{\mathrm{x}} = \Delta H_{\mathrm{x=0}} + 2\alpha\mathrm{x} + 3\beta\mathrm{x}^2 + 4\gamma\mathrm{x}^3
\end{equation}

\begin{figure}[h]
\centering
\includegraphics[width=\columnwidth]{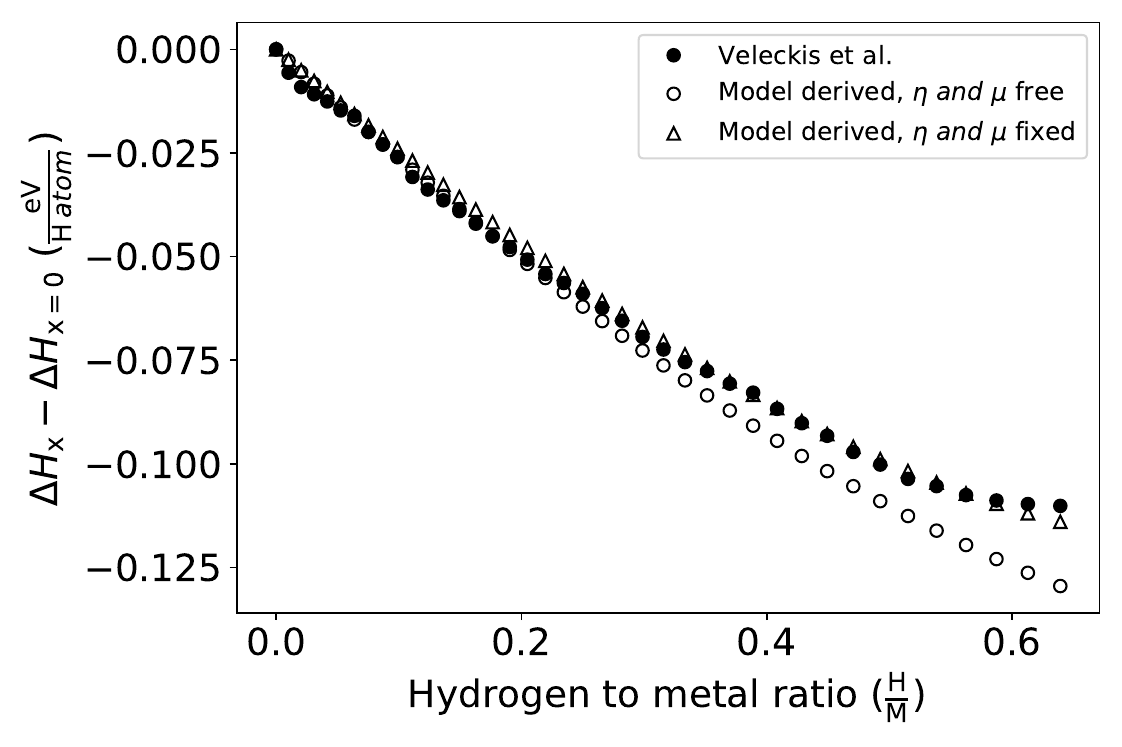} 
\caption{Variation of $\Delta H_{\mathrm{x}}^{\circ} - \Delta H_{\mathrm{x=0}}^{\circ}$ with $\mathrm{x}$}
\label{fig:4.16}       
\end{figure}

In Eq. \ref{eq:relative-partial-molal-enthalpy-temp-independent}, $\Delta H_{\mathrm{x=0}}$ is the $\Delta H_{\mathrm{x}}$ value at infinite dilution. In Fig. \ref{fig:4.16}, we plot the $\Delta H_{\mathrm{x}} - \Delta H_{\mathrm{x=0}}$ values derived from the $\alpha$, $\beta$, $\gamma$ values presented in Table \ref{tab:4.2} corresponding to the two limits of the model. 
We also plot the values reported by Veleckis et al.\ \cite{veleckisThermodynamicPropertiesSystems1969a} for comparison. We note that in contrast to the quality of the isotherms reported earlier, the $\Delta H_{\mathrm{x}}$ values derived through the parameters corresponding to the first limit of the model ($\eta = \boldsymbol{\eta}$ and $\mu = \boldsymbol{\mu}$) are in better agreement with the values reported by Veleckis et al.\ \cite{veleckisThermodynamicPropertiesSystems1969a}, compared to the values derived from the parameters corresponding to the second limit of the model ($\eta$ and $\mu$ are free). Since the second limit of the model yields a better fit to the absorption isotherms, we expect the values for $\Delta H_{\mathrm{x}}- \Delta H_{\mathrm{x=0}}$ derived from the second limit of the phenomenological model to be more accurate compared to the corresponding values derived from the first limit of the model or the values reported by Veleckis et al.\ \cite{veleckisThermodynamicPropertiesSystems1969a}. \\

The relative partial molal entropy of hydrogen at a hydrogen-to-metal ratio of x is given by 

\begin{equation}
\label{eq:relative-partial-molal-entropy}
\Delta S_{\mathrm{x}} = \bar{S}_{\mathrm{H}}^{\mathrm{NbH}} - \frac{1}{2} S_{\mathrm{H_2}}^{\circ}
\end{equation}

Without temperature dependence, $\Delta S_{\mathrm{x}}$ can be expressed as in Eq. \ref{eq:S_H^NbH-model-derived-temp-independent}. Therefore we get

\begin{equation}
\label{eq:relative-partial-molal-entropy-temp-independent}
\Delta S_{\mathrm{x}} = - k_B\mathrm{\log}\biggr[\frac{\mathrm{x}}{R_{mean}-\mathrm{x}}\biggr] - \frac{1}{2} S_{\mathrm{H_2}}^{\circ}
\end{equation}

\begin{figure}[h]
\centering
\includegraphics[width=\columnwidth]{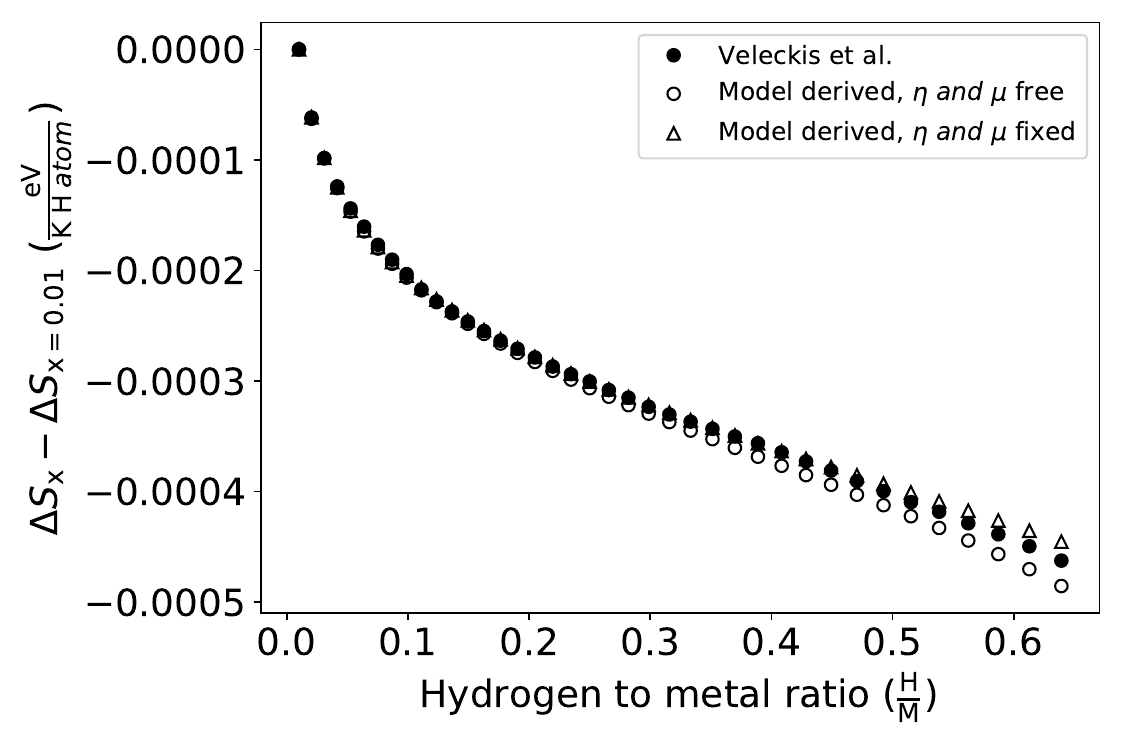}
\caption{Variation of $\Delta S_{\mathrm{x}}^{\circ} - \Delta S_{\mathrm{x=0.01}}^{\circ}$ with $\mathrm{x}$}
\label{fig:4.17}       
\end{figure}

Using the $R_{mean}$ values from Table \ref{tab:4.2} in Eq. \ref{eq:relative-partial-molal-entropy-temp-independent}, we can calculate $\Delta S_{\mathrm{x}}$ corresponding to the two limits of the model. We reference the entropy of hydrogen dissolution ($\Delta S_{\mathrm{x}}$) values with respect to the infinite dilution limit values ($\Delta S_{\mathrm{x=0.01}}$) and present them in Fig. \ref{fig:4.17}. Note that infinite dilution limit for $\Delta S_{\mathrm{x}}$ is calculated at $\mathrm{x=0.01}$ as opposed to $\mathrm{x=0}$, to not violate the domain of the log function in 
Eq. \ref{eq:relative-partial-molal-entropy-temp-independent}. In Fig. \ref{fig:4.17}, we plot the referenced entropy values derived from the parameters corresponding to the model's two limits. We also plot the corresponding values reported by Veleckis et al.\ \cite{veleckisThermodynamicPropertiesSystems1969a} for comparison. We note that the entropy trend derived from the two limits of the model and that reported by Veleckis et al.\ \cite{veleckisThermodynamicPropertiesSystems1969a} are very similar. At low concentrations, the values for the entropy of hydrogen dissolution reported by Veleckis et al.\ \cite{veleckisThermodynamicPropertiesSystems1969a} are identical to those derived from the two limits of our model. At higher concentrations, the entropy values reported by Veleckis et al.\ \cite{veleckisThermodynamicPropertiesSystems1969a} are sandwiched by the entropy values derived from our model's two limits. \\

\begin{figure}[h]
\centering
\includegraphics[width=\columnwidth]{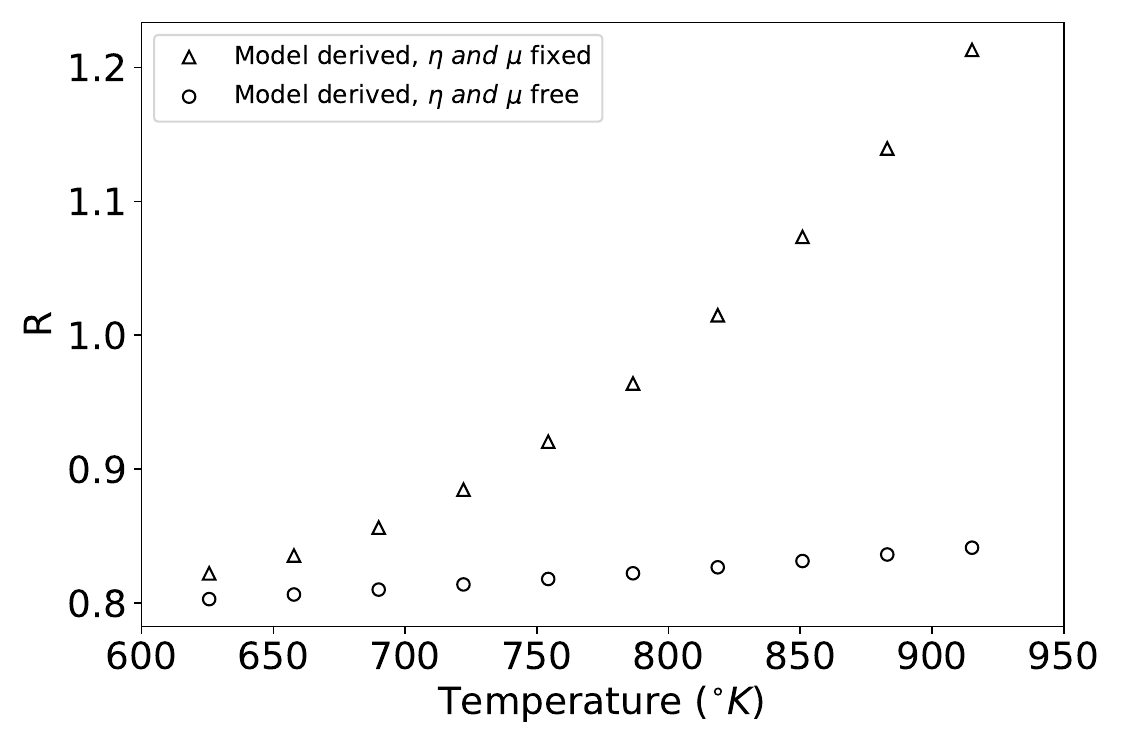}
\caption{Variation of $R$ With temperature}
\label{fig:4.18}       
\end{figure}

The temperature variation of $\Delta S_{\mathrm{x}}$ can be understood by looking at the variation of $R$ with temperature. We calculate this temperature dependence of $R$ using Eq. \ref{eq:R_temp_dep} and plot it in Fig. \ref{fig:4.18}. \\

The first limit of the model yields a set of $R$ values ranging from 0.8 to 1.28 in the temperature range of interest. The second limit of the model yields a set of $R$ values ranging from 0.80 to 0.85. For the $\mathrm{Nb-H}$ system, Kuji et al. \cite{kujiKuji1984Thermodynamic1984} report $R$ can take a maximum value up to $R_{max}$, which varies according to how many nearest neighboring sites of an occupied site are excluded for occupation. In Table \ref{tab:4.3}, we show these $R_{max}$ values and how many nearest neighboring sites are blocked corresponding to each $R_{max}$ value.

\begin{table}[h]
\caption{Limiting compositions $R_{max}$ when sites to the $(n - 1)^{th}$ nearest neighbor of an occupied site are excluded from occupation. Adapted from Kuji et al. \cite{kujiKuji1984Thermodynamic1984}.}
\begin{tabular}{cccc} 
\hline
$n^{th}$ nearest neighbor & $R_{max}$ \\
\hline

& \\

1 & 6\\

& \\

2 & 3 \\

& \\

3 & 1.5\\

& \\

4 & 1.25 \\

& \\

6 & 1 \\

& \\

7 & 0.75 \\

& \\

\hline
\end{tabular}
\label{tab:4.3}
\end{table}

The exclusion in site occupancy is due to the short-range repulsive interactions between the interstitial hydrogen atoms in niobium discussed in our previous work. The entropic contribution to the Gibbs free energy dominates at higher temperatures, leading to a less restrictive exclusion of site occupancy. Hence, the observed increase in $R$ values with temperatures reported in Fig \ref{fig:4.18}. \\

The $R_{max}$ values reported in Table \ref{tab:4.3} put the $R$ values in Fig \ref{fig:4.18} into perspective and give us an idea of how many sites are likely to be available for occupation over the temperature range of interest here (626.65 - 944.15 K). According to the $R$ values corresponding to the first limit of the model, at 626.65 K, we can expect up to the fifth nearest neighboring sites of an occupied site to be excluded for occupation. Whereas at 944.15 K, the blocking only extends up to the second nearest neighbor. According to the $R$ values corresponding to the second limit of the model, we can expect the blocking to extend up to the fifth nearest neighbor in the entire temperature range. \\

\section{Validation of the Phenomenological Model}\label{sec5}

In all the fitting procedures performed thus far, we have fit our phenomenological model to the complete data set from Veleckis et al.\ \cite{veleckisThermodynamicPropertiesSystems1969a} and evaluated the parameters that represent the best fit. From these parameters, we extracted valuable inferences about the interactions between interstitial hydrogen atoms and derived thermodynamic quantities like the enthalpy and entropy of hydrogen dissolution in niobium. \\ 

However, we have yet to validate the phenomenological model on data that was not used to fit the model. Here, the k-fold cross-validation method will be used to evaluate the phenomenological model's performance on unseen data. We chose k to be 5 for a reasonably sized data set for testing. Therefore, we divide the absorption isotherm data from Veleckis et al.\ \cite{veleckisThermodynamicPropertiesSystems1969a} into five groups of equal size ($\sim$41 data points per group) and populate them with randomly selected data points. Then, for each unique group, holding the data points in that group as the testing set, we train the phenomenological model on all the data points from the remaining (k-1) folds and note down the parameters corresponding to the best fit to the training data. We then evaluate the model's performance on the testing data set and calculate the RMSE value. We repeat this procedure for each unique group, obtain the average value of the model's parameters, and evaluate the aggregate cross-validation RMSE. We let $\eta$ and $\mu$ be free parameters when performing the 5-fold cross-validation procedure. \\

In Table \ref{tab:4.4}, we present the mean value of the model parameters and aggregate cross-validation RMSE value obtained via the 5-fold cross-validation procedure. For comparison, we also present the values corresponding to the fit to the entire data set setting $\eta$ and $\mu$ as free parameters (Same values as reported in the second column of Table \ref{tab:4.2}). \\

\begin{table}[h]
\caption{A comparison of the mean value of model parameters and the aggregate cross-validation RMSE obtained via the 5-fold cross-validation procedure to model parameters and RMSE value obtained by fitting the complete absorption isotherm data from Veleckis et al.\ \cite{veleckisThermodynamicPropertiesSystems1969a} to the phenomenological model. $\eta$ and $\mu$ were set as free parameters in both cases.}

\begin{tabular}{cccc} 
\hline
Method & 5-fold cross-validation  & Fit to the entire data set \\
\hline

& & \\

$H\ (\frac{\mathrm{eV}}{\mathrm{H}\ atom})$ & -0.36 & -0.36 \\

& & \\

$\alpha\ (\frac{\mathrm{eV}}{\mathrm{(H/Nb)}^2})$ & -0.14& -0.14  \\

& & \\

$\beta\ (\frac{\mathrm{eV}}{\mathrm{(H/Nb)}^3})$ & 0.03 & 0.03 \\ 

& & \\

$\gamma\ (\frac{\mathrm{eV}}{\mathrm{(H/Nb)}^4})$ & 0.01 & 0.01 \\

& & \\

$R_0$ & 0.78 & 0.78 \\

& & \\

$R_1$ & -$2.0 \times 10^{-5}$ & -$1.8 \times 10^{-5}$ \\

& & \\

$R_2$ & $1.0 \times 10^{-7}$ & $9.8 \times 10^{-8}$ \\

& & \\

$R_{mean}$ & 0.82 & 0.82 \\

& & \\
 
$\eta$ & 6.69 & 6.69 \\ 

& & \\

$\mu$ & $1.3 \times 10^{-3}$ &  $1.3 \times 10^{-4}$ \\

& & \\

RMSE & 0.94 & 0.85 \\

& & \\
\hline
\end{tabular}
\label{tab:4.4}
\end{table}

From Table \ref{tab:4.2}, we note that the aggregate cross-validation RMSE is within 10\% of the RMSE value corresponding to the fit to the entire data set. The closeness of the RMSE values is a good indicator of the robustness of the model and indicates that the model's performance on unseen data is satisfactory. Additionally, the mean value of the parameters from the 5-fold cross-validation is almost identical to the values of the parameters corresponding to the fit to the entire data set, further indicating that the model is robust. \\

\section{Conclusion}\label{sec6}

The phenomenological model developed in this work describes the thermodynamics of hydrogen absorption in niobium. Veleckis et al. \cite{veleckisThermodynamicPropertiesSystems1969a} were limited to the scope of the data analysis carried out in their work primarily due to the lack of computational and data analytic tools that are available today. Revisiting these old experiments with modern tools shows that their data have high accuracy and pass the test of time. \\

The phenomenological model has far fewer parameters than the one used by Veleckis et al. \cite{veleckisThermodynamicPropertiesSystems1969a} while also improving the physical description of this system by capturing the temperature dependence of enthalpy and entropy of hydrogen in the metal and gas phases. Furthermore, this model lets you extract more information from the experimental data than reported in the original work. More specifically, the model gives physical insights into the nature of interactions between interstitial hydrogen atoms in niobium and the number of interstitial sites available for hydrogen occupation at a given temperature. We also derived the enthalpy and entropy of hydrogen dissolution in niobium from these model parameters and found them to be in good agreement with the corresponding values reported by Veleckis et al. \cite{veleckisThermodynamicPropertiesSystems1969a} \\

The developments and findings from this work also lay the foundation to study the thermodynamics of this system using first principle methods. For instance, since our analysis shows that several interstitial sites are available at a given temperature for hydrogen occupation, we can anticipate that sampling the interstitial configurational space is critical when calculating the enthalpy of hydrogen dissolution in niobium. \\

Finally, the model described here could be applied to describe the absorption isotherms of other metal hydrogen systems and can thus impact the application of metals in hydrogen separation and transport applications. \\

\backmatter

\section{Supplementary information}

This article has a supplementary file containing supporting information. 

\section{Acknowledgments}

We gratefully acknowledge the financial support from the Center for Negative Carbon Emissions at Arizona State University. 

\section{Conflict of interest}

The authors declare that they have no conflict of interest.

\section{Authors' contributions}
A.R. developed the theoretical formalism, analyzed the data, and wrote the manuscript. Both H.Z. and K.S.L. supervised this work and
contributed to the final version of the manuscript.


\bibliography{Pheno_model}


\begin{thebibliography}{17}
\ifx \bisbn   \undefined \def \bisbn  #1{ISBN #1}\fi
\ifx \binits  \undefined \def \binits#1{#1}\fi
\ifx \bauthor  \undefined \def \bauthor#1{#1}\fi
\ifx \batitle  \undefined \def \batitle#1{#1}\fi
\ifx \bjtitle  \undefined \def \bjtitle#1{#1}\fi
\ifx \bvolume  \undefined \def \bvolume#1{\textbf{#1}}\fi
\ifx \byear  \undefined \def \byear#1{#1}\fi
\ifx \bissue  \undefined \def \bissue#1{#1}\fi
\ifx \bfpage  \undefined \def \bfpage#1{#1}\fi
\ifx \blpage  \undefined \def \blpage #1{#1}\fi
\ifx \burl  \undefined \def \burl#1{\textsf{#1}}\fi
\ifx \doiurl  \undefined \def \doiurl#1{\url{https://doi.org/#1}}\fi
\ifx \betal  \undefined \def \betal{\textit{et al.}}\fi
\ifx \binstitute  \undefined \def \binstitute#1{#1}\fi
\ifx \binstitutionaled  \undefined \def \binstitutionaled#1{#1}\fi
\ifx \bctitle  \undefined \def \bctitle#1{#1}\fi
\ifx \beditor  \undefined \def \beditor#1{#1}\fi
\ifx \bpublisher  \undefined \def \bpublisher#1{#1}\fi
\ifx \bbtitle  \undefined \def \bbtitle#1{#1}\fi
\ifx \bedition  \undefined \def \bedition#1{#1}\fi
\ifx \bseriesno  \undefined \def \bseriesno#1{#1}\fi
\ifx \blocation  \undefined \def \blocation#1{#1}\fi
\ifx \bsertitle  \undefined \def \bsertitle#1{#1}\fi
\ifx \bsnm \undefined \def \bsnm#1{#1}\fi
\ifx \bsuffix \undefined \def \bsuffix#1{#1}\fi
\ifx \bparticle \undefined \def \bparticle#1{#1}\fi
\ifx \barticle \undefined \def \barticle#1{#1}\fi
\bibcommenthead
\ifx \bconfdate \undefined \def \bconfdate #1{#1}\fi
\ifx \botherref \undefined \def \botherref #1{#1}\fi
\ifx \url \undefined \def \url#1{\textsf{#1}}\fi
\ifx \bchapter \undefined \def \bchapter#1{#1}\fi
\ifx \bbook \undefined \def \bbook#1{#1}\fi
\ifx \bcomment \undefined \def \bcomment#1{#1}\fi
\ifx \oauthor \undefined \def \oauthor#1{#1}\fi
\ifx \citeauthoryear \undefined \def \citeauthoryear#1{#1}\fi
\ifx \endbibitem  \undefined \def \endbibitem {}\fi
\ifx \bconflocation  \undefined \def \bconflocation#1{#1}\fi
\ifx \arxivurl  \undefined \def \arxivurl#1{\textsf{#1}}\fi
\csname PreBibitemsHook\endcsname

\bibitem[\protect\citeauthoryear{Veleckis and
  Edwards}{1969}]{veleckisThermodynamicPropertiesSystems1969a}
\begin{barticle}
\bauthor{\bsnm{Veleckis}, \binits{E.}},
\bauthor{\bsnm{Edwards}, \binits{R.K.}}:
\batitle{Thermodynamic properties in the systems vanadium-hydrogen,
  niobium-hydrogen, and tantalum-hydrogen}.
\bjtitle{The Journal of Physical Chemistry}
\bvolume{73}(\bissue{3}),
\bfpage{683}--\blpage{692}
(\byear{1969})
\end{barticle}
\endbibitem

\bibitem[\protect\citeauthoryear{Albrecht
  et~al.}{1959}]{albrechtReactionsNiobiumHydrogen1959a}
\begin{barticle}
\bauthor{\bsnm{Albrecht}, \binits{W.M.}},
\bauthor{\bsnm{Goode}, \binits{W.D.}},
\bauthor{\bsnm{Mallett}, \binits{M.W.}}:
\batitle{Reactions in the {Niobium}‐{Hydrogen} {System}}.
\bjtitle{Journal of The Electrochemical Society}
\bvolume{106}(\bissue{11}),
\bfpage{981}--\blpage{986}
(\byear{1959})
\end{barticle}
\endbibitem

\bibitem[\protect\citeauthoryear{Kuji and
  Oates}{1984}]{kujiKuji1984Thermodynamic1984}
\begin{barticle}
\bauthor{\bsnm{Kuji}, \binits{T.}},
\bauthor{\bsnm{Oates}, \binits{W.A.}}:
\batitle{Kuji (1984) {Thermodynamic} properties of {Nb}-{H} alloys {I}: {The}
  $\alpha$ phase}.
\bjtitle{Journal of the Less Common Metals}
\bvolume{102}(\bissue{2}),
\bfpage{251}--\blpage{260}
(\byear{1984})
\end{barticle}
\endbibitem

\bibitem[\protect\citeauthoryear{Komjathy}{1960}]{komjathyNiobiumhydrogenSystem1960}
\begin{barticle}
\bauthor{\bsnm{Komjathy}, \binits{S.}}:
\batitle{The niobium-hydrogen system}.
\bjtitle{Journal of the Less Common Metals}
\bvolume{2}(\bissue{6}),
\bfpage{466}--\blpage{480}
(\byear{1960})
\end{barticle}
\endbibitem

\bibitem[\protect\citeauthoryear{{Schaumann G.}
  et~al.}{1970}]{schaumanng.Schaumann1970Nb2006}
\begin{barticle}
\bauthor{\bsnm{{Schaumann G.}}},
\bauthor{\bsnm{{Völki J.}}},
\bauthor{\bsnm{{Alefeld G.}}}:
\batitle{Schaumann (1970) {H} and {D} in {Nb}, {Va}, {Ta}}.
\bjtitle{physica status solidi (b)}
\bvolume{42}(\bissue{1}),
\bfpage{401}--\blpage{413}
(\byear{1970})
\end{barticle}
\endbibitem

\bibitem[\protect\citeauthoryear{Wipf and
  Alefeld}{1974}]{wipfDiffusionCoefficientHeat}
\begin{barticle}
\bauthor{\bsnm{Wipf}, \binits{H.}},
\bauthor{\bsnm{Alefeld}, \binits{G.}}:
\batitle{Diffusion coefficient and heat of transport of {H} and {D} in niobium
  below room temperature}.
\bjtitle{physica status solidi (a)}
\bvolume{23}(\bissue{1}),
\bfpage{175}--\blpage{186}
(\byear{1974})
\end{barticle}
\endbibitem

\bibitem[\protect\citeauthoryear{Peterson and
  Jensen}{1978}]{petersonPeterson1978Nb1978}
\begin{barticle}
\bauthor{\bsnm{Peterson}, \binits{D.T.}},
\bauthor{\bsnm{Jensen}, \binits{C.L.}}:
\batitle{Peterson (1978) {H},{D} in {Nb}/{Va}}.
\bjtitle{Metallurgical Transactions A}
\bvolume{9}(\bissue{11}),
\bfpage{1673}--\blpage{1677}
(\byear{1978})
\end{barticle}
\endbibitem

\bibitem[\protect\citeauthoryear{Erckmann and
  Wipf}{1976}]{erckmannErckman1976Nb1976}
\begin{barticle}
\bauthor{\bsnm{Erckmann}, \binits{V.}},
\bauthor{\bsnm{Wipf}, \binits{H.}}:
\batitle{Erckman (1976) {H}/{D} in {Nb},{V}, {Ta}}.
\bjtitle{Physical Review Letters}
\bvolume{37}(\bissue{6}),
\bfpage{341}--\blpage{344}
(\byear{1976})
\end{barticle}
\endbibitem

\bibitem[\protect\citeauthoryear{Brouwer and
  Griessen}{1989}]{brouwerBrouwer1989Nb1989}
\begin{barticle}
\bauthor{\bsnm{Brouwer}, \binits{R.C.}},
\bauthor{\bsnm{Griessen}, \binits{R.}}:
\batitle{Brouwer (1989) {H} in {Nb}}.
\bjtitle{Physical Review Letters}
\bvolume{62}(\bissue{15}),
\bfpage{1760}--\blpage{1763}
(\byear{1989})
\end{barticle}
\endbibitem

\bibitem[\protect\citeauthoryear{Wipf}{1976}]{wipfWipf1975Metals1976}
\begin{barticle}
\bauthor{\bsnm{Wipf}, \binits{H.}}:
\batitle{Wipf (1975) {H} in {Metals}}.
\bjtitle{Journal of the Less Common Metals}
\bvolume{49},
\bfpage{291}--\blpage{307}
(\byear{1976})
\end{barticle}
\endbibitem

\bibitem[\protect\citeauthoryear{Ramachandran
  et~al.}{2020}]{ramachandranProbingInteractionsInterstitial2020}
\begin{barticle}
\bauthor{\bsnm{Ramachandran}, \binits{A.}},
\bauthor{\bsnm{Zhuang}, \binits{H.}},
\bauthor{\bsnm{Lackner}, \binits{K.S.}}:
\batitle{Probing the interactions between interstitial hydrogen atoms in
  niobium through density functional theory calculations}.
\bjtitle{Materials Today Communications}
\bvolume{25},
\bfpage{101415}
(\byear{2020})
\doiurl{10.1016/j.mtcomm.2020.101415} .
Accessed 2021-02-27
\end{barticle}
\endbibitem

\bibitem[\protect\citeauthoryear{Rohatgi}{2020}]{Rohatgi2020}
\begin{botherref}
\oauthor{\bsnm{Rohatgi}, \binits{A.}}:
Webplotdigitizer: Version 4.4
(2020).
\url{https://automeris.io/WebPlotDigitizer}
\end{botherref}
\endbibitem

\bibitem[\protect\citeauthoryear{SciPy.org}{2021}]{scipyint60:online}
\begin{botherref}
\oauthor{\bsnm{SciPy.org}}:
Univariate spline interpolation (scipy.interpolate.UnivariateSpline), SciPy
  v1.7.1 {Manual}
(2021).
\url{https://docs.scipy.org/doc/scipy/reference/generated/scipy.interpolate.UnivariateSpline.html}
\end{botherref}
\endbibitem

\bibitem[\protect\citeauthoryear{Lee
  et~al.}{2015}]{leeUnderstandingDeviationsHydrogen2015a}
\begin{barticle}
\bauthor{\bsnm{Lee}, \binits{K.}},
\bauthor{\bsnm{Yuan}, \binits{M.}},
\bauthor{\bsnm{Wilcox}, \binits{J.}}:
\batitle{Understanding {Deviations} in {Hydrogen} {Solubility} {Predictions} in
  {Transition} {Metals} through {First}-{Principles} {Calculations}}.
\bjtitle{The Journal of Physical Chemistry C}
\bvolume{119}(\bissue{34}),
\bfpage{19642}--\blpage{19653}
(\byear{2015})
\doiurl{10.1021/acs.jpcc.5b05469} .
\bcomment{Publisher: American Chemical Society}.
Accessed 2021-06-09
\end{barticle}
\endbibitem

\bibitem[\protect\citeauthoryear{Chase~Jr and Tables}{1998}]{chase1998data}
\begin{barticle}
\bauthor{\bsnm{Chase~Jr}, \binits{M.W.}},
\bauthor{\bsnm{Tables}, \binits{N.-J.T.}}:
\batitle{Data reported in nist standard reference database 69, june 2005
  release: Nist chemistry webbook}.
\bjtitle{J. Phys. Chem. Ref. Data, Monograph}
\bvolume{9},
\bfpage{1}--\blpage{1951}
(\byear{1998})
\end{barticle}
\endbibitem

\bibitem[\protect\citeauthoryear{Boggs and
  Rogers}{1990}]{boggsOrthogonalDistanceRegression1990}
\begin{botherref}
\oauthor{\bsnm{Boggs}, \binits{P.T.}},
\oauthor{\bsnm{Rogers}, \binits{J.E.}}:
Orthogonal {Distance} {Regression},
16
(1990)
\end{botherref}
\endbibitem

\bibitem[\protect\citeauthoryear{SciPy.org}{2021}]{OrthogonalDistanceRegression2021}
\begin{botherref}
\oauthor{\bsnm{SciPy.org}}:
Orthogonal distance regression (scipy.odr), SciPy v1.7.1 {Manual}
(2021).
\url{https://docs.scipy.org/doc/scipy/reference/odr.html}
\end{botherref}
\endbibitem

\end{thebibliography}

\end{document}


\title{\ \ \ \ \ \ \ \ \ \ \ \ \ Supporting Information \\ \\
A phenomenological model for interstitial hydrogen absorption in niobium}

\author*[1]{\fnm{Arvind} \sur{Ramachandran}}\email{arvindramachandran0522@gmail.com}

\author[2]{\fnm{Houlong} \sur{Zhuang}}\email{hzhuang7@asu.edu}

\author[1]{\fnm{Klaus} \sur{S. Lackner}}\email{Klaus.Lackner@asu.edu}

\affil*[1]{School of Sustainable Engineering and the Built Environment, Arizona State University, Tempe, USA}

\affil[2]{School for Engineering of Matter, Transport and Energy, Arizona State University, Tempe, USA}

\maketitle

\section{Reproducing Veleckis et al.'s Data Analysis}

Here we outline the details of repeating the fitting procedure used by Veleckis et al. \cite{veleckisThermodynamicPropertiesSystems1969a} to derive relative partial molal enthalpy and the relative partial molal entropy at varying hydrogen concentrations. \\

We considered a set of hydrogen-to-metal ratios, $\{\mathrm{x}_i\}$, corresponding to the mole fraction values, $\{\mathrm{y}_i\}$, chosen by Veleckis et al.\ \cite{veleckisThermodynamicPropertiesSystems1969a} and estimated the set of equilibrium pressures at each $\mathrm{x}_i$ and the various temperatures corresponding to the different isotherms. For each $\mathrm{x}_i$, we fit the set of $\{\mathrm{\log}(P_i^{\frac{1}{2}})\}$ values to an expression $A+\frac{B}{T}$ following Eq. \ref{eq:veleckis_eqbm_condition}. \\

\begin{equation}
\tag{S1}
\label{eq:veleckis_eqbm_condition}
\mathrm{\log}(P^{\frac{1}{2}}) = A + \frac{B}{T}
\end{equation}

In Fig. \ref{fig:S1}, we plot five such fits at $\mathrm{x}_i = 0.02, 0.11, 0.22, 0.35, 0.52$ along with the original $\mathrm{\log}(P(\mathrm{x_i}, T)^{\frac{1}{2}})$ values. From Fig. \ref{fig:S1} it is evident that the $log(P(\mathrm{x_i}, T)^{\frac{1}{2}})$ indeed can be expressed as $A+\frac{B}{T}$ at different hydrogen concentrations with high accuracies. We find the $R^2$ values for the fit to every set of $\{\mathrm{\log}(P_i^{\frac{1}{2}})\}$ values to be greater than 0.99, with the mean value of all fits exceeding 0.999. \\

\renewcommand{\thefigure}{S1:}
\begin{figure}[h]
\includegraphics[width=\columnwidth]{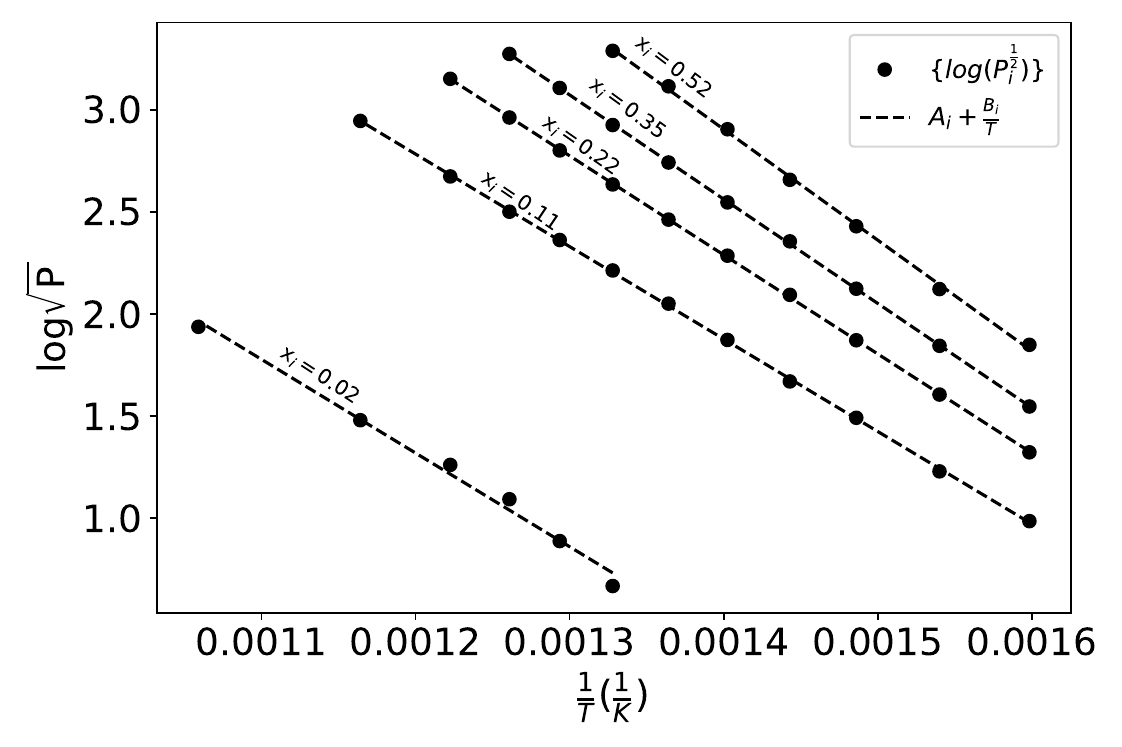}
\caption{Goodness of fitting $\mathrm{\log}P^{\frac{1}{2}}$ to the expression $\mathrm{A + \frac{B}{T}}$}
\label{fig:S1}      
\end{figure}

We use the set of A and B values at various concentrations to calculate the relative partial molal enthalpy and entropy according to equations \ref{eq:relative_enthalpy_B_relation} and \ref{eq:relative_entropy_A_relation}, respectively. In Fig. \ref{fig:S2} and Fig. \ref{fig:S3}, we plot the calculated relative partial molal enthalpy and relative partial molal entropy at varying hydrogen-to-metal ratios, respectively. We also plot the relative partial molal enthalpy and relative partial molal entropy reported by Veleckis et al.\ \cite{veleckisThermodynamicPropertiesSystems1969a} for comparison. As shown in Fig. \ref{fig:S2} and Fig. \ref{fig:S3}, the relative partial molal enthalpy and relative partial molal entropy values derived by repeating Veleckis et al.'s \cite{veleckisThermodynamicPropertiesSystems1969a} data analysis procedure are in very close agreement with the corresponding values reported by Veleckis et al. \cite{veleckisThermodynamicPropertiesSystems1969a}. \\

\begin{equation}
\tag{S2}
\label{eq:relative_enthalpy_B_relation}
\bar{H}_{\mathrm{H}}^{\mathrm{NbH}} - \frac{1}{2} H_{\mathrm{H_2}}^{\circ} = k_BB
\end{equation}

\begin{equation}
\tag{S3}
\label{eq:relative_entropy_A_relation}
\bar{S}_{\mathrm{H}}^{\mathrm{NbH}} - \frac{1}{2} S_{\mathrm{H_2}}^{\circ} = k_B(\mathrm{\log}(P_0^{\frac{1}{2}})-A)
\end{equation}

\renewcommand{\thefigure}{S2:}
\begin{figure}[H]
\includegraphics[width=\columnwidth]{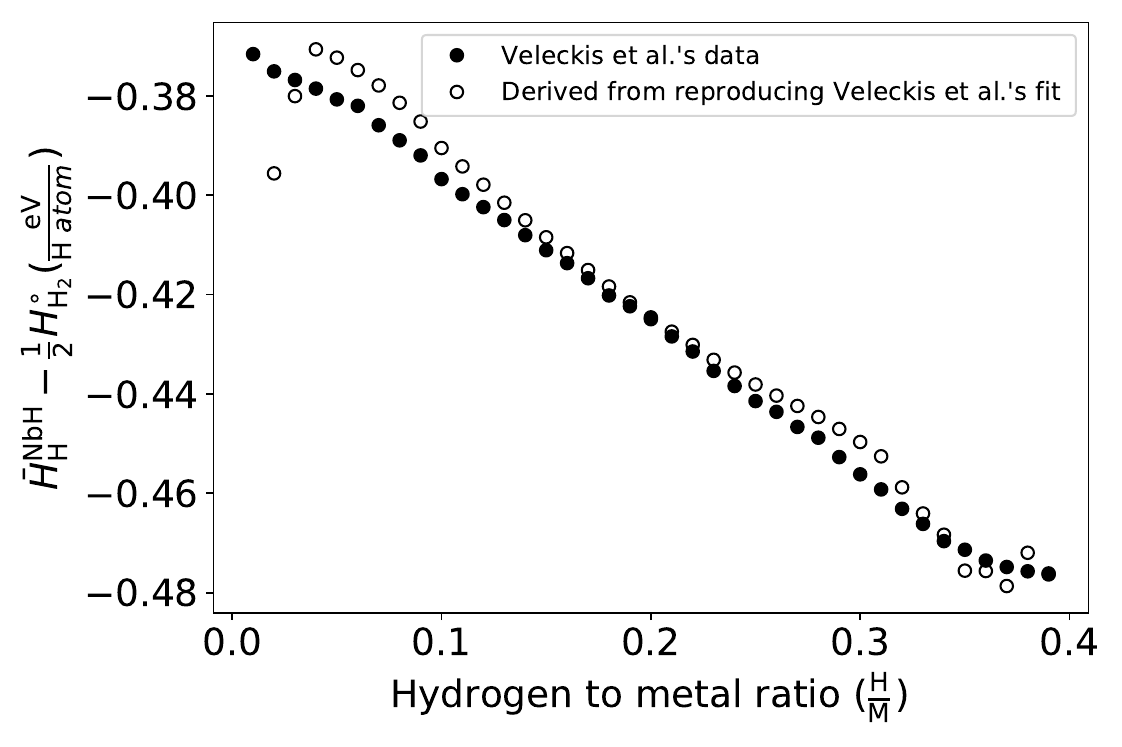}
\caption{Comparing the $\bar{H}_{\mathrm{H}}^{\mathrm{NbH}} - \frac{1}{2} H_{\mathrm{H_2}}^{\circ}$ values derived from repeating the fitting procedure by Veleckis et al. \cite{veleckisThermodynamicPropertiesSystems1969a} with the $\bar{H}_{\mathrm{H}}^{\mathrm{NbH}} - \frac{1}{2} H_{\mathrm{H_2}}^{\circ}$ values reported by Veleckis et al.\ \cite{veleckisThermodynamicPropertiesSystems1969a}}
\label{fig:S2}      
\end{figure}

In Fig. \ref{fig:S2} and Fig. \ref{fig:S3}, the slight differences between the values reported by Veleckis et al.\ \cite{veleckisThermodynamicPropertiesSystems1969a} and the values derived through repeating their analysis could be due to a few reasons. While plotting the values reported by Veleckis et al.\ \cite{veleckisThermodynamicPropertiesSystems1969a} we do not account for the error bars for $\bar{H}_{\mathrm{H}}^{\mathrm{NbH}} - \frac{1}{2} H_{\mathrm{H_2}}^{\circ}$ and $\bar{S}_{\mathrm{H}}^{\mathrm{NbH}} - \frac{1}{2} S_{\mathrm{H_2}}^{\circ}$ reported by Veleckis et al.\ \cite{veleckisThermodynamicPropertiesSystems1969a}. Furthermore, some errors could be introduced in digitizing the data, and slight differences in connecting the points using a smoothing spline as opposed to the smooth curves drawn by Veleckis et al.\ \cite{veleckisThermodynamicPropertiesSystems1969a}. Nevertheless, more importantly, being able to reproduce the values for $\bar{H}_{\mathrm{H}}^{\mathrm{NbH}} - \frac{1}{2} H_{\mathrm{H_2}}^{\circ}$ and $\bar{S}_{\mathrm{H}}^{\mathrm{NbH}} - \frac{1}{2} S_{\mathrm{H_2}}^{\circ}$ reported by Veleckis et al.\ \cite{veleckisThermodynamicPropertiesSystems1969a} with reasonable accuracy indicates that we understand the physical model and fitting procedure used by Veleckis et al.\ \cite{veleckisThermodynamicPropertiesSystems1969a}. \\

\renewcommand{\thefigure}{S3:}
\begin{figure}[H]
\includegraphics[width=\columnwidth]{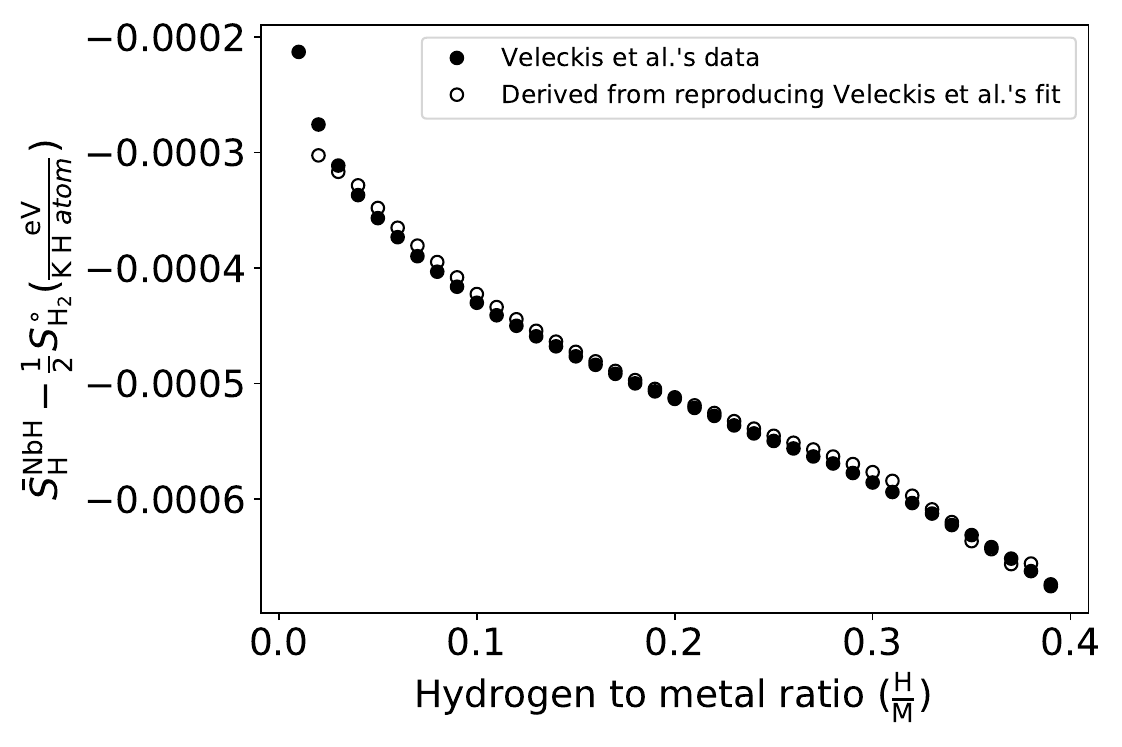}
\caption{Comparing the $\bar{S}_{\mathrm{H}}^{\mathrm{NbH}} - \frac{1}{2} S_{\mathrm{H_2}}^{\circ}$ values derived from repeating the fitting procedure by Veleckis et al.\ \cite{veleckisThermodynamicPropertiesSystems1969a} with the $\bar{S}_{\mathrm{H}}^{\mathrm{NbH}} - \frac{1}{2} S_{\mathrm{H_2}}^{\circ}$ values reported by Veleckis et al.\ \cite{veleckisThermodynamicPropertiesSystems1969a}}
\label{fig:S3}      
\end{figure}

The A and B values at various concentrations were also used to calculate the pressure values at various concentrations and temperatures. The $\mathrm{\sqrt{Pressure}}$ values are plotted in Fig. \ref{fig:S4}. The digitized pressure data and the smoothing spline fit from Fig. 2 in the main article are also plotted for comparison. As expected, the pressure values calculated from the set of A and B closely align with the original data points reported by Veleckis et al.\ \cite{veleckisThermodynamicPropertiesSystems1969a}, and the smoothing spline fit to the original data points. \\

\renewcommand{\thefigure}{S4:}
\begin{figure}[H]
\includegraphics[width=\columnwidth]{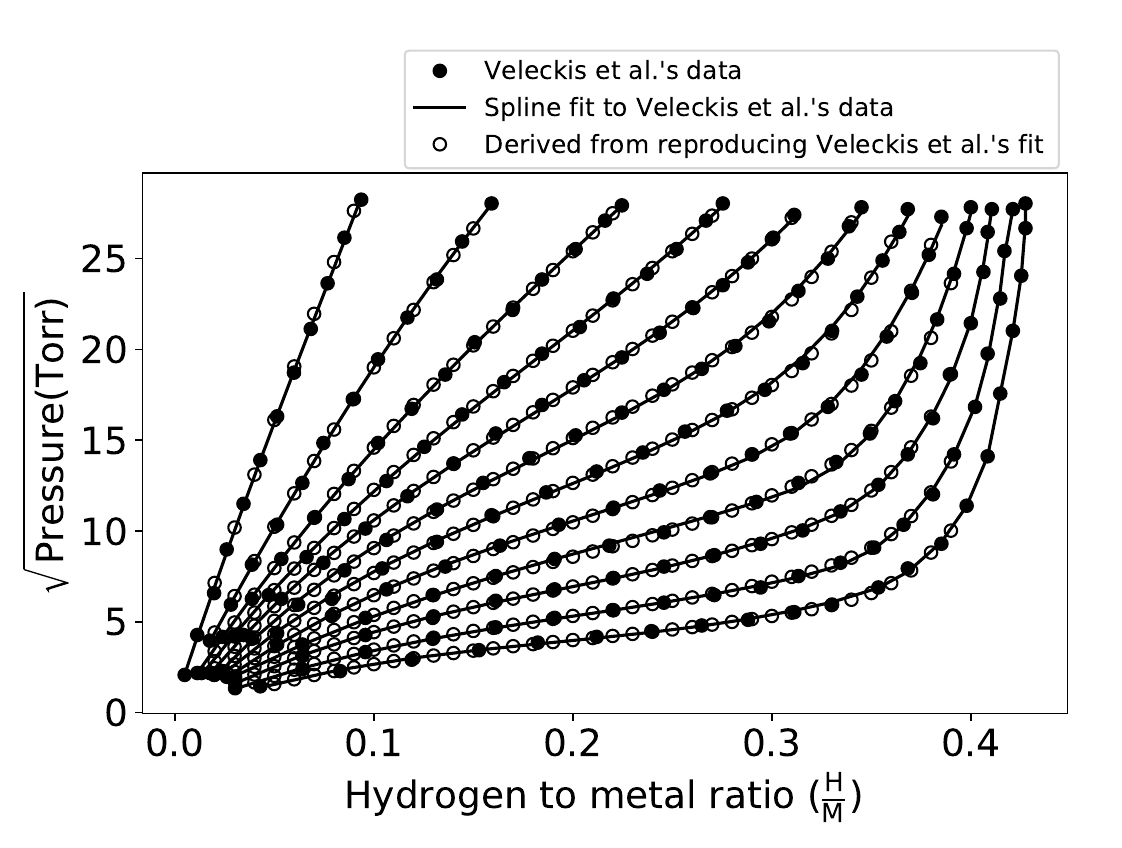}
\caption{Comparing the equilibrium pressure values derived from the set of $\{A_i\}$ and $\{B_i\}$ values with Veleckis et al.'s \cite{veleckisThermodynamicPropertiesSystems1969a} original data and the smoothing spline fit to their data}
\label{fig:S4}      
\end{figure}

